\documentclass[11pt]{JHEP3}
\usepackage{amsmath, amssymb, bm,cite}
\title{The canonical transformation and massive CSW vertices for MHV-SQCD}
\author{ Tim~R.~Morris and
  Zhiguang~Xiao\\
  School of Physics and Astronomy,  University of Southampton\\
  Highfield, Southampton, SO17 1BJ, U.K.\\
  E-mails: 
\email{T.R.Morris@soton.ac.uk}, \email{z.g.xiao@phys.soton.ac.uk} }

\preprint{SHEP 08-39}


\abstract{The similarity of massive CSW scalar vertices  and quark
vertices can be understood using a kind of light-cone SUSY
transformation presented in this paper. We also show that the
canonical transformation generating the MHV-SQCD lagrangian, can be
fixed by applying this  light-cone SUSY transformation to the
canonical transformation for MHV-QCD obtained in paper
arxiv:0805.0239. Most of the massive CSW vertices for SQCD can also be
pinned down in this way. }

\begin{document}
\section{Introduction} 
The striking simplicity of MHV amplitudes has been an inspiration for
many recent developments in gauge theory. It was first proposed by
Parke and Taylor \cite{Parke:1986gb} and then proved using recursion
methods by Berends and Giele \cite{Berends:1988zn}. After about
fifteen years, inspired by twistor string theory, Cachazo, Svr\v cek
and Witten discovered that tree-level amplitudes with arbitrary
helicity configuration can be constructed using MHV amplitudes
continued off shell in a particular way and connected using scalar
propagators \cite{Cachazo:2004kj}. This proposal reduces the
complexity of the amplitude computation dramatically.  CSW rules were
then extended to include quarks and to supersymmetric theories
\cite{Wu:2004fb,Wu:2004jxa,Georgiou:2004wu,Georgiou:2004by}, and also
were extended to include Higgs \cite{Dixon:2004za,Badger:2004ty} and
electroweak gauge bosons \cite{Bern:2004ba}.  There are also some
applications of CSW rules in one loop calculations (such as in \cite{
Brandhuber:2004yw, Bedford:2004py, Bedford:2004nh, Brandhuber:2005kd,
Quigley:2004pw, Brandhuber:2006bf, Brandhuber:2007vm}). 

As well as the proof given in \cite{Britto:2005fq}, a particularly
direct proof of these rules was found in \cite{Risager:2005vk} by
generalising the idea from BCFW recursion relations
\cite{Britto:2004ap}.  But despite a preliminary attempt to derive the
MHV lagrangian \cite{Gorsky:2005sf}, a full and clear understanding of
the origin of CSW rules from quantum field theory was not presented
until Mansfield's paper \cite{Mansfield:2005yd}. In his paper,
Mansfield proposed a framework for deriving the MHV lagrangian using a
certain canonical transformation applied to the light-cone lagrangian:
CSW rules then followed from the vertices in this MHV lagrangian.  The
concrete canonical transformation was obtained in \cite{Ettle:2006bw}
and later was extended in \cite{Ettle:2007qc} to give a prescription
for dimensional regularization of the MHV lagrangian. It was also made
clear that the previously missing pieces of amplitudes in the CSW
prescription can be recovered by so called `completion vertices',
which come from the field transformation. Another approach to obtain
the MHV lagrangian is to use the twistor Yang-Mills theory
\cite{Boels:2007qn,Boels:2006ir} by formulating the gauge theory on
twistor space and then fixing a particular gauge.

In \cite{Boels:2007pj,Boels:2008ef}, a canonical transformation was
used to generate the massive CSW vertices involving massive scalars,
and it was shown that these can also be obtained from the twistor
Yang-Mills approach. A full canonical transformation for QCD including
a quark field was constructed in a recent paper \cite{Ettle:2008ey},
in which CSW vertices for massive quarks are also presented.  The same
results have been reproduced from the twistor Yang-Mills approach
\cite{Boels:2008du} (although formulated in a different convention).  

The similarity between the massive CSW scalar vertices in
\cite{Boels:2007pj,Boels:2008ef} and massive quark vertices in
\cite{Ettle:2008ey} suggests a supersymmetric relation between these
two vertices. In \cite{Schwinn:2008fm}, the author uses a massive
version of the supersymmetric Ward identity (massive SWI)
\cite{Schwinn:2006ca} to understand this similarity. Since massive SWI
can only be used on amplitudes, one should choose a suitable reference
momentum of the massive quark to make the amplitudes involving a
massive quark-antiquark pair be proportional only to the corresponding
massive CSW vertices. Then the massive SWI can be used to relate them
to amplitudes involving massive scalars, hence relating these two
kinds of CSW vertices. Since the derivation of the massive SWI makes
use of the explicit solution of massive Dirac equation, it also
depends on the reference momenta of the solution.  However, at the
level of the MHV lagrangian, one would not expect that a
supersymmetric relation between vertices should depend on the on-shell
solution of the Dirac equation. In the present paper, a kind of
light-cone supersymmetric transformation at the lagrangian level is
presented to relate these two kinds of vertices directly. It is based
on the observation that after the gauge fixing and integrating out of
the non-dynamical fields, there is some supersymmetry left in the
lagrangian.  Moreover, the complete canonical transformation for the
MHV lagrangian for supersymmetric QCD (MHV-SQCD) can be obtained by
using this supersymmetry transformation on the results already
obtained for MHV-QCD in \cite{Ettle:2008ey}.  Besides the relation
between massive CSW vertices for massive quarks and scalars, other
relations among massive CSW vertices for MHV-SQCD can also be obtained
in this way. As a result, most massive CSW vertices can be fixed using
these relations and the results from MHV-QCD.  Another kind of
light-cone supersymmetry is used in \cite{Feng:2006yy} in discussing
MHV lagrangian  for ${\cal N}=4$ gauge theory. 
  
The paper is organised as follows: In section \ref{sect:convention},
we explain the conventions and notation used in this paper, which is a
little different from \cite{Ettle:2008ey}. In section
\ref{sect:lc-SQCD-trans}, we derive the light-cone lagrangian and
provide the light-cone SUSY transformation.  In section
\ref{sect:canonical-trans-SQCD}, we obtain all of the canonical
transformation for MHV-SQCD using the light-cone SUSY transformation
and the canonical transformation for MHV-QCD.  In section
\ref{sect:CSW-massiveV-SQCD}, we derive the massive CSW vertices for
MHV-SQCD. Section \ref{sect:conclusion} contains the conclusions and
discussion.  Since we will use the canonical transformation and
massive CSW vertices for MHV-QCD from \cite{Ettle:2008ey} in our
derivation, we summarise them in appendix \ref{sect:sum-canonical-QCD}
and \ref{sect:mass-V-QCD}. In appendix \ref{sect:lc-SQCD-L}, we give
the full light-cone lagrangian for SQCD. In appendix
\ref{sect:sum-canonical-SQCD} we summarise our results for the
canonical transformation for SQCD.

\section{Preliminary}
\label{sect:convention}
\subsection{Light-cone co-ordinates}
The light-cone co-ordinates are defined as:
\begin{equation}
       x^0 = \tfrac1{\sqrt 2}(t-x^3), \quad
x^{\bar 0} = \tfrac1{\sqrt 2}(t+x^3), \quad
         z = \tfrac1{\sqrt 2}(x^1+ix^2), \quad
    \bar z = \tfrac1{\sqrt 2}(x^1-ix^2).
\label{eq:lc-4d-coords}
\end{equation}
We employ a compact notation writing $( p_0,p_{\bar0},p_z,p_{\bar
z})\equiv (\check p, \hat p,\tilde p, \bar p)$ and for momenta
labelled by a number we write that number with decorations to denote
the components of the momentum: $(\check p_n, \hat p_n,\tilde p_n,
\bar p_n)\equiv (\check n, \hat n,\tilde n, \bar n)$. For component
$\tilde p$, we will omit the tilde in case it causes no confusion.  In
this notation, the Lorentz invariant reads
\begin{equation}
A\cdot B=\hat A \check B+\check A \hat B- A \bar B -\bar A B\,.
\end{equation} 
We also make extensive use of the bilinears:
\begin{eqnarray}
(i\:j)=\hat k_i \tilde k_j-\hat k_j\tilde k_i\,,\quad 
\{i\:j\}=\hat k_i \bar k_j -\hat k_j\bar k_i\,.
\end{eqnarray}

\subsection{Spinor conventions}
We use the Weyl representation of the Dirac matrices:
\begin{eqnarray}
&&\gamma^\mu=\left(
\begin{array}{cc} 
0&\sigma^\mu_{\alpha\dot\alpha}
\cr \bar\sigma^{\mu,\dot\alpha \alpha} &0
\end{array}
\right)\,,
\\
&&\hspace{-3cm}
\sigma^\mu_{\alpha\dot\alpha}
=(1,\vec\sigma^i)\,,
\quad
\bar\sigma^{\mu,\dot\alpha \alpha} =(1,-\vec\sigma^i)
\,,\quad 
\sigma^{\mu\nu}=\frac 1 4 (\sigma^\mu\bar
\sigma^\nu-\sigma^\nu\bar\sigma^\mu)\,.
\end{eqnarray}
The massless Weyl spinors are solutions of Dirac equation 
$p_\mu\sigma^\mu_{\alpha\dot\alpha} \bar\chi^{\dot\alpha}=0$ and 
$p_\mu\bar\sigma^{\mu,\dot\alpha\alpha} \chi_\alpha=0$: 
\begin{eqnarray}
\chi_\alpha=2^{1/4}(\sqrt{\hat p},\bar p/\sqrt {\hat p})^{\rm T}
\,,\quad 
\bar \chi^{\dot \alpha}=2^{1/4}(p/\sqrt {\hat p},-\sqrt{\hat p})^{\rm
T}
\\
\chi^\alpha=2^{1/4}(\bar p/\sqrt {\hat p},-\sqrt{\hat p})^{\rm T}
\,,\quad 
\bar \chi_{\dot \alpha}=2^{1/4}(\sqrt{\hat p},p/\sqrt {\hat p})^{\rm
T}
\end{eqnarray}
where $\chi^\alpha=\varepsilon^{\alpha \beta}\chi_\alpha$, 
$\bar\chi_{\dot\alpha}=\varepsilon_{\dot\alpha \dot\beta}
\bar\chi^{\dot\beta}$ and 
$\varepsilon^{12}=\varepsilon^{\dot1\dot2}=-\varepsilon_{1,2}=-\varepsilon_{\dot1\dot2}=1$.

The usual spinor bilinear products can be recast into our convention  
\begin{eqnarray}
\langle i\:j\rangle=
\bar\chi_{\dot\alpha}(k_i)\bar\chi^{\dot\alpha}(k_j)=\sqrt2{(i\:j)\over
\sqrt{\hat \imath \:\hat \jmath}}\,,
\\
{[} i\:j{]}=
\chi^{\alpha}(k_i)\chi_{ \alpha}(k_j)
=-\sqrt2{\{i\:j\}\over \sqrt{\hat \imath\:\hat \jmath}}\,.
\end{eqnarray}
\subsection{SQCD Lagrangian and component fields}
We will be working with the supersymmetric QCD Lagrangian with two chiral superfields:
\begin{eqnarray}
L_{\rm SQCD}&=&\int\!\!{\rm d}^3x\, \Big[\Phi_1^\dagger e^{-2iV^*} \Phi_1 +\Phi_2^\dagger
e^{-2iV}\Phi_2\Big]\Big|_{\theta^4} -\frac 1 {2 g^2}{\rm tr}\Big[W^\alpha
W_\alpha\Big|_{\theta^2}+h.c.\Big]
\nonumber\\
&&\qquad+m(\Phi_1\Phi_2\Big|_{\theta^2}+h.c.)
\end{eqnarray}
where $\Phi_2$ and $\Phi_1$ are the chiral superfields in the $N$ and
$\bar N$ representation of color $SU(N)$ group respectively and
$W_\alpha$ is the field strength spinor superfield of the gauge field.
$(\phi_1, \psi_1,F_1)$, $(\phi_2,\psi_2,F_2)$ are component fields of
$\Phi_1$ and $\Phi_2$, and ($\Lambda_\alpha$, ${\cal F}_{\mu\nu}$,$D$)
are component fields of $W_\alpha$, ${\cal A}_\mu$ being the corresponding
component gauge field in $V$:  
\begin{eqnarray}
\Phi(y)&=&\phi(y)+\sqrt2\theta \psi (y)+\theta^2 F(y)  
\,,\\
W_\alpha(y)&=&
\Lambda_\alpha(y)+\theta^\beta\Big(-D(y)\varepsilon_{\beta\alpha}+ 
i(\sigma^{\mu\nu})_{\beta}{}^{\gamma}\varepsilon_{\alpha\gamma}{\cal
F}_{\mu\nu}(y)\Big)
\nonumber \\
&&+\theta^2(i\sigma^\mu_{\alpha\dot\alpha}{\cal
D}_\mu\bar\Lambda^{\dot \alpha}(y))\,,
\end{eqnarray}
where 
\begin{equation}
y^\mu=x^\mu-i\theta\sigma^\mu\bar \theta\,.
\end{equation}
The gauge field strength is defined as
\begin{equation}
  \label{eq:gauge-conv}
  {\cal F}_{\mu\nu} = [{\cal D}_\mu, {\cal D}_\nu], \quad
  {\cal D}_\mu = \partial_\mu + {\cal A}_\mu, \quad
  {\cal A}_\mu = - \frac{ig}{\sqrt 2}  A^a_\mu \tau^a\,,\quad
T^a=-i\frac{\tau^a}{\sqrt 2},
\end{equation}
where the normalization of color matrix $\tau^a$ is:
\begin{equation} 
[\tau^a, \tau^b ] = i \sqrt 2 f^{abc} \tau^c, 
\quad 
{\rm tr} (\tau^a \tau^b) = \delta^{ab}.
\end{equation}
${\cal D}_\mu$ for gluino is defined as ${\cal D}_\mu \Lambda=
\partial_\mu\Lambda + [{\cal A}_\mu,\Lambda]$. 

Before gauge fixing, the SUSY
transformations for component fields are:
\begin{eqnarray}
\delta {\cal A}_\mu&=&\eta \sigma_\mu \bar \Lambda+\Lambda \sigma_\mu
\bar \eta
\\
\delta \Lambda_\alpha&=&D\eta_\alpha
-i(\sigma^{\mu\nu})_\alpha{}^{\beta}\eta_{\beta}{\cal F}_{\mu\nu}
\nonumber \\
\delta D&=&-i\eta \sigma^\mu {\cal D}_\mu \bar\Lambda
-i\bar \eta \bar \sigma^\mu {\cal D}_\mu \Lambda 
\nonumber \\
\delta  \phi_{1,2}&=& \sqrt2 \eta^\alpha(\psi_{1,2})_\alpha\,,
\nonumber \\
\delta(\psi_{1,2})_\alpha&=&\sqrt2(\eta_\alpha F_{1,2}-i[ \sigma^\mu\bar \eta]_\alpha
{\cal D}_\mu\phi_{1,2})
\nonumber\\
\delta F_{1,2}&=&-\sqrt2 i\bar \eta\bar \sigma^\mu {\cal
D}_\mu\psi_{1,2}-2\bar
\eta\bar \lambda\phi_{1,2}\,.
\label{eq:full-susy}
\end{eqnarray}
After integrating out the auxilary fields $F$ and $D$ we
find the SQCD lagrangian in component fields:
\begin{eqnarray}
L_{\rm SQCD}&=&({\cal D}_\mu \phi_1)^\dagger {\cal D}^\mu
\phi_1+({\cal D}_\mu
\phi_2)^\dagger {\cal D}^\mu \phi_2-m^2 ( \phi_1^\dagger \phi_1+
\phi_2^\dagger \phi_2)
\nonumber \\
&&+i(\psi_1^\alpha {\cal D}\!\!\!\!/_{\alpha\dot\alpha}\bar
\psi_1^{\dot\alpha}+\bar\psi_2^{\dot\alpha} {\cal D}\!\!\!\!/_{\dot\alpha
\alpha}\bar \psi_2^{\alpha})-m (\psi_1^\alpha
\psi_{2\alpha}+ \bar\psi_{1,\dot\alpha}\bar\psi_{2}^{\dot\alpha})
\nonumber \\
&&-\sqrt2i(\phi_1^{\rm T} \bar \Lambda
\bar\psi_1+\psi_1\Lambda\phi_1^*)+ \sqrt2i(\bar\psi_2 \bar \Lambda
\phi_2+\phi_2^\dagger\Lambda\psi_2)
\nonumber \\
&&-\frac 2{g^2} {\rm tr}\bigg[-\frac 14
{\cal F}_{\mu\nu}{\cal F}^{\mu\nu}+\frac i 2(\Lambda^\alpha
{\cal D}\!\!\!\!/_{\alpha\dot\alpha}\bar \Lambda^{\dot \alpha}+\bar \Lambda^{\dot
\alpha}{\cal D}\!\!\!\!/_{\dot\alpha\alpha}\Lambda^\alpha)\bigg]
\nonumber \\
&&+g^2(- \phi_1^{\rm T} T^a \phi_1 ^* +\phi_2^\dagger T^a \phi_2)^2
\,.
\label{eq:SUSY-L}
\end{eqnarray}
We can combine the two Weyl spinors to be a Dirac spinor
\begin{eqnarray}
\Psi=(\psi_{2,\alpha},\bar\psi_1^{\dot\alpha})^{\rm T}
\end{eqnarray}
and like in \cite{Ettle:2008ey}, we denote the components of the spinor as  
\begin{equation}
  \Psi = (\alpha^+, \beta^+, \beta^-, \alpha^-)^{\rm T}
  \quad\text{and}\quad
  \bar\Psi = (\bar\beta^+, \bar\alpha^+, \bar\alpha^-, \bar\beta^-)\,,
\end{equation}
where the $\pm$ superscripts denote the physical helicity of the
outgoing particles for massless theory and $\bar
\alpha^\pm=(\alpha^{\mp})^*$.  For the gluino, we denote the
components of Weyl spinor $\Lambda$ and $\bar \Lambda$ as
\begin{equation}
\Lambda_\alpha=(\Lambda,T)\,,
\quad 
\bar\Lambda^{\dot\alpha}=(\bar T,-\bar\Lambda)\,.
\end{equation}
We also denote $\phi_1(\phi_1^*)$ as $\bar\phi^+(\bar\phi^-)$ and
$\phi_2(\phi_2^*)$ as $\phi^+(\phi^-)$. In this way, we will see later
that  the superscript $\pm$ of $\alpha$, $\bar \alpha$ will be the
same as  their superpartners'.  As an abuse of the nomenclature we
will also call  $\pm$ superscripts of scalars as
plus(minus)-chirality. 

Fields in momentum space are defined as the Fourier
transformation:
\begin{eqnarray}
  f(x)&=&\int {{\rm d}\hat q \,{\rm d}\tilde q\,{\rm d}\bar q \over (2\pi)^3} f(\vec q)
 e^{i (\hat q x^{\bar 0}+ \tilde q x^z+ \bar q x^{\bar z})}
\end{eqnarray}
and we use numbered subscripts to denote the momenta labelled with
numbers of the fields:
\begin{equation}
f_1=f(\vec p_1)\,,\quad f_{\bar 1}=f(-\vec p_1)\,.
\end{equation}
We also use a short-hand notation for the momentum integral product
\begin{equation}
  \int_{1\cdots n} = \prod_{k=1}^n \frac 1{(2\pi)^3}
  \int d\hat k \: dk \: d\bar k.
\end{equation}

\section{Light-cone SQCD lagrangian and light-cone SUSY
transformations}
\label{sect:lc-SQCD-trans}
We start with SQCD lagrangian (\ref{eq:SUSY-L}). As in
\cite{Mansfield:2005yd,Ettle:2006bw,Ettle:2007qc,Ettle:2008ey}, we
quantise the theory on the constant $x^0$ surface $\Sigma$ with a
normal vector $\mu=(1,0,0,1)/\sqrt2$ in Minkowski co-ordinates  and
choose the light-cone gauge $\hat {\cal A}=0$.  Then we find out that
the dynamical fields are $\bar {\cal A}$, ${\cal A}$, $\bar
\alpha^{\pm}$, $\alpha^{\pm}$, $\Lambda$, $\bar\Lambda$, $\bar
\phi^\pm$, $\phi^\pm$, whilst $\check{\cal A}$, $T$, $\bar T$,
$\beta^\pm$, $\bar\beta^\pm$ can be integrated out. After this, we can
group the terms in the light-cone SQCD lagrangian according to their
chirality configuration and their field content:
\begin{eqnarray}
L_{\rm LCSQCD}&=& L^{+-}_A+ L^{+-}_{\Lambda}+ L^{+-}_{\phi}+ L^{+-}_{\alpha}
\nonumber \\
&&+ L^{++-}_{A}+ L^{++-}_{\Lambda A}+ L^{++-}_{\alpha A}+ L^{++-}_{\phi
A}+ L^{++-}_{\alpha\Lambda \phi}
\nonumber \\
&&+ L^{--+}_{A}+ L^{--+}_{\Lambda A}+ L^{--+}_{\alpha A}+ L^{--+}_{\phi
A}+ L^{--+}_{\alpha\Lambda \phi}
\nonumber \\
&&+ L^{--++}_{A}+L^{--++}_{\Lambda}+  L^{--++}_{\Lambda
A}+L^{--++}_{\alpha}+L^{--++}_{\phi}+  L^{--++}_{\alpha\phi}
\nonumber \\&&
+  L^{--++}_{\alpha A}+ L^{--++}_{\phi
A}+ L^{--++}_{\alpha\Lambda}+ L^{--++}_{\phi\Lambda} +
L^{--++}_{\alpha\phi \Lambda A}
\nonumber \\&&
+ L^{+-}_{m,\alpha}+ L^{+-}_{m,\phi}
+ L^{+-+}_{m,\alpha A}+ L^{+-+}_{m,\phi \Lambda \alpha}
+ L^{-+-}_{m,\alpha A}+ L^{-+-}_{m,\phi \Lambda \alpha}\,,
\end{eqnarray}
in which the superscripts differentiate their chirality configurations
and the subscripts denote their field content. Subscript $m$ in the
last line labels massive terms. $L^{+-}_{m,\cdots}$ terms are
proportional to $m^2$, whilst $L^{+-+}_{m,\cdots}$ and
$L^{-+-}_{m,\cdots}$ are proportional to $m$.  The full expressions
for each term are summarised in appendix \ref{sect:lc-SQCD-L}. The
$L^{+-}_\alpha$, $L^{+-}_{m,\alpha}$ and $L^{+-}_\phi$,
$L^{+-}_{m,\phi}$ terms result in  massive propagators for $\alpha$ and
$\phi$:
\begin{eqnarray}
\langle \alpha^\pm\bar\alpha^{\mp}\rangle=\frac { i\,\sqrt 2 \hat
p}{p^2 -m^2}\,, \quad
\langle \phi^-\phi^{+}\rangle=\langle\bar
\phi^-\bar\phi^{+}\rangle=\frac { i }{p^2 -m^2}\,.
\end{eqnarray}
Since we have fixed the light-cone gauge, the full SUSY transformation
will not preserve the gauge, but there is a subgroup of the full SUSY
transfromation which  leaves the light-cone gauge invariant. Also,
since only the dynamical fields are left in the light-cone SQCD
lagrangian, the remaining supersymmetry can only involve these fields.
If one restricts the SUSY transformation parameters $\eta_\alpha$ and
$\bar\eta^{\dot \alpha}$ in (\ref{eq:full-susy}) to be
\begin{eqnarray}
\eta_\alpha =(0,\eta) \,,\quad \bar\eta^{\dot\alpha}=(\bar\eta,0)\,,
\end{eqnarray}
one finds that this subgroup of the full  SUSY transformation does
indeed preserve the gauge condition and the space of dynamical fields.
To be specific, these transformations for dynamical fields are
\begin{eqnarray}
\delta \bar \phi^+=-\sqrt2 \eta \bar \alpha^+ \,,
&\quad&
\delta \bar \phi^-=\sqrt2 \bar \eta \alpha^- \,,
\label{eq:lc-susy-1}\\
\delta \phi^+=\sqrt2 \eta  \alpha^+ \,,
&\quad&
\delta \phi^-=-\sqrt2 \bar \eta \bar\alpha^- \,,
\label{eq:lc-susy-2}\\
\delta \bar \alpha^-=2i\eta \hat \partial \phi^-\,,
&\quad&
\delta \alpha^+=-2i\bar\eta \hat \partial \phi^+\,,
\label{eq:lc-susy-3}\\
\delta \alpha^-=-2i\eta \hat \partial \bar\phi^-\,,
&\quad&
\delta \bar \alpha^+=2i\bar\eta \hat \partial \bar \phi^+\,,
\label{eq:lc-susy-4}\\
\delta \Lambda=2i\eta \hat \partial {\cal A} \,,
&\quad&
\delta \bar \Lambda=-2i\bar\eta \hat \partial \bar {\cal A}\,,
\label{eq:lc-susy-5}\\
\delta \bar {\cal A}=\sqrt2\eta \bar \Lambda \,,
&\quad&
\delta {\cal A}=-\sqrt2\bar\eta \Lambda \,.
\label{eq:lc-susy-6}
\end{eqnarray}
The auxiliary fields $F$, $D$ and the nonlinear terms also
automatically disappear in this transformation.  We will call this
SUSY transformation the light-cone SUSY transformation. Furthermore,
we can group the dynamical fields into pairs:
\begin{eqnarray}
\{\alpha^+,\phi^+\} \,, \{\alpha^-,\bar \phi^-\}\,,
\{\bar \alpha^-,\phi^-\} \,, \{\bar \alpha^+,\bar \phi^+\}\,,\{\Lambda,
{\cal A}\},\{\bar \Lambda, \bar{\cal A}\}
\end{eqnarray}
in which each pair of the dynamical fields generate an  invariant
sub-space under this light-cone SUSY transformation. As a result, the
terms in the light-cone SQCD can also be grouped into SUSY invariant
pieces:
\begin{eqnarray}
&&\{L^{+-}_A+ L^{+-}_{\Lambda}\}
\,,\quad
\{L^{+-}_{\phi}+
L^{+-}_{\alpha}\}\,,
\nonumber\\
&& \{L^{++-}_{A}+ L^{++-}_{\Lambda A}\}
\,, \quad
\{ L^{++-}_{\alpha A}+ L^{++-}_{\phi
A}+ L^{++-}_{\alpha\Lambda \phi}\}\,,
\nonumber\\
&&\{L^{--+}_{A}+ L^{--+}_{\Lambda A}\}
\,, \quad
\{L^{--+}_{\alpha A}+ L^{--+}_{\phi
A}+ L^{--+}_{\alpha\Lambda \phi} \}\,,
\nonumber\\
&&\{ L^{--++}_{A}+L^{--++}_{\Lambda}+L^{--++}_{\Lambda
A}\}
\,,\quad
\{ L^{--++}_{\alpha}+L^{--++}_{\phi}+  L^{--++}_{\alpha\phi} \}\,,
\nonumber\\&&
 \{ L^{--++}_{\alpha A}+ L^{--++}_{\phi
A}+ L^{--++}_{\alpha\Lambda}+ L^{--++}_{\phi\Lambda} +
L^{--++}_{\alpha\phi \Lambda A}\}\,,
\nonumber\\&&
\{ L^{+-}_{m,\alpha}+ L^{+-}_{m,\phi} \}
\,,\quad
\{ L^{+-+}_{m,\alpha A}+ L^{+-+}_{m,\phi \Lambda \alpha}\}
\,,\quad
\{L^{-+-}_{m,\alpha A}+ L^{-+-}_{m,\phi \Lambda \alpha}\}\,.
\label{eq:L-inv}
\end{eqnarray}
These are not the smallest pieces that are invariant under these
transformations. For example, one can also separate the
$L^{+-}_{\phi}+ L^{+-}_{\alpha}$ into two parts containing
$\{\phi^\pm,\alpha^+,\bar\alpha^-\}$ and
$\{\bar\phi^\pm,\alpha^-,\bar\alpha^+\}$ respectively which are
separately invariant under the light-cone SUSY transformations.
However, if one term in one curly bracket of (\ref{eq:L-inv}) appears
as a whole, to be closed under these transformations one must add the
other terms in this curly bracket.

\section{Canonical transformations for the MHV-SQCD}
\label{sect:canonical-trans-SQCD}
From previous work \cite{Ettle:2008ey}, we have already got the
canonical transformations for $\alpha^\pm$, $\bar \alpha^\pm$, $\cal A$
and $\bar {\cal A}$ for the MHV lagrangian of QCD. And now we also
have the light-cone SUSY transformation
(\ref{eq:lc-susy-1})--(\ref{eq:lc-susy-6}). In this section, we will
fix the canonical transformations for the SQCD using these two
results.

The canonical field pairs before and after the canonical
transformation for SQCD are (up to some irrelavent normalizations):
\begin{eqnarray}
&&\{ {\cal A},\hat\partial\bar {\cal A}\}\to\{ {\cal B},\hat\partial\bar {\cal B}\}
\,,\quad
\{\Lambda,\bar \Lambda\}\to\{\Pi,\bar \Pi\}
\,,\quad
\nonumber\\
&&
\{\alpha^\pm,\bar \alpha^\mp\}\to\{\xi^\pm,\bar \xi^\mp\}
\,,\quad
\nonumber\\
&&
\{\phi^-,\hat \partial\phi^+\}\to \{\varphi^-,\hat \partial\varphi^+\}
\,,\quad
\{\bar \phi^-,\hat \partial\bar\phi^+\}\to\{\bar \varphi^-,\hat
\partial\bar\varphi^+\}.
\end{eqnarray}

As in \cite{Mansfield:2005yd,Ettle:2006bw,Ettle:2008ey}, the canonical
transformation should transform the massless non-MHV terms to kinetic
terms:
\begin{eqnarray}
&&L^{+-}_A+ L^{+-}_{\Lambda}+ L^{+-}_{\phi}+ L^{+-}_{\alpha}
+ L^{++-}_{A}+ L^{++-}_{\Lambda A}
\nonumber\\
&&\quad\quad\quad+ L^{++-}_{\alpha A}+ L^{++-}_{\phi
A}+ L^{++-}_{\alpha\Lambda \phi}
=L^{+-}_B+ L^{+-}_{\Pi}+
L^{+-}_{\varphi}+ L^{+-}_{\xi}
\,.\label{eq:canonical-eq}
\end{eqnarray}
The canonical transformation can be represented as a power 
expansion of the old fields in terms of the new fields. Like equations
(\ref{eq:lc-susy-1})--(\ref{eq:lc-susy-6}), we expect the SUSY
transformation for the new fields to be linear, so that it is
satisfied at each order of the expansion.  Since to the leading order
of the transformation expansion, the old fields are the same as the
new fields, we make the natural assumption that the new fields
actually satisfy the same light-cone SUSY transformations as the
corresponding old fields:
\begin{eqnarray}
\delta \bar \varphi^+=-\sqrt2 \eta \bar \xi^+ \,,
&\quad&
\delta \bar \varphi^-=\sqrt2 \bar \eta \xi^- \,,
\label{eq:lc-susy-1-new}\\
\delta \varphi^+=\sqrt2 \eta  \xi^+ \,,
&\quad&
\delta \varphi^-=-\sqrt2 \bar \eta \bar\xi^- \,,
\label{eq:lc-susy-2-new}\\
\delta \xi^-=-2i\eta \hat \partial \bar\varphi^-\,,
&\quad&
\delta \bar \xi^+=2i\bar\eta \hat \partial \bar \varphi^+\,,
\label{eq:lc-susy-3-new}\\
\delta \bar \xi^-=2i\eta \hat \partial \varphi^-\,,
&\quad&
\delta \xi^+=-2i\bar\eta \hat \partial \varphi^+\,,
\label{eq:lc-susy-4-new}\\
\delta \Pi=2i\eta \hat \partial {\cal B} \,,
&\quad&
\delta \bar \Pi=-2i\bar\eta \hat \partial \bar {\cal B}\,,
\label{eq:lc-susy-5-new}\\
\delta \bar {\cal B}=\sqrt2\eta \bar \Pi\,,
&\quad&
\delta {\cal B}=-\sqrt2\bar\eta \Pi \,.
\label{eq:lc-susy-6-new}
\end{eqnarray}
As further support for this assumption, we note that this implies that
both the left hand side and the right hand side of
(\ref{eq:canonical-eq}) are then closed under the light-cone SUSY
transformation. 

Recall that the equation defining the canonical transformation
equation for MHV-QCD is only part of equation (\ref{eq:canonical-eq}):
\begin{eqnarray} 
&&L^{+-}_A+ L^{+-}_{\alpha} + L^{++-}_{A}+
L^{++-}_{\alpha A} =L^{+-}_B+ L^{+-}_{\xi}\,.
\label{eq:canonical-eq-QCD} 
\end{eqnarray} 
Since each term on the left hand side of this equation is in the first
four curly brackets separately in (\ref{eq:L-inv}) and also each term
on the right hand side is in the first two curly brackets separately,
the minimal supersymmetric extension of this equation just gives us
back the whole equation (\ref{eq:canonical-eq}). Therefore, if we can
find a canonical transformation that transforms correctly under the
SUSY transformation and is also consistent with what we already have
in MHV-QCD,  we will have solved the canonical transformation equation
(\ref{eq:canonical-eq}).  To be specific, we can separate the
transformation expansion for fields $\{ {\cal A}, \bar {\cal A},
\alpha^\pm, \bar \alpha^\pm\}$ into QCD pieces which involve only QCD
fields $\{ {\cal B}, \bar {\cal B}, \xi^\pm, \bar \xi^\pm\}$, and new
pieces which contain new supersymmetric fields $
\{\varphi^\pm,\bar\varphi^{\pm},\Pi,\bar \Pi\}$ as well as QCD fields:
\begin{eqnarray}
X=X^{\rm QCD}[{\cal B}, \bar {\cal B}, \xi^\pm, \bar \xi^\pm]+X^{\rm
New}[{\cal B}, \bar {\cal B}, \xi^\pm, \bar
\xi^\pm,\varphi^\pm,\bar\varphi^{\pm},\Pi,\bar \Pi]\,,\quad X\in \{ {\cal A}, \bar {\cal A},
\alpha^\pm, \bar \alpha^\pm\},
\nonumber \\
\end{eqnarray}
in which either each expansion term in $X^{\rm New}$ contains at least
one field from $ \{\varphi^\pm,\bar\varphi^{\pm},\Pi,\bar \Pi\}$ or
$X^{\rm New}=0$. At the same time, one can see that the expansion of
$\{\Lambda, \bar \Lambda, \phi^\pm,\bar \phi^\pm\} $ should not
contain pieces which only have QCD fields:
\begin{eqnarray}
{\cal X}={\cal X}[{\cal B}, \bar {\cal B}, \xi^\pm, \bar
\xi^\pm,\varphi^\pm,\bar\varphi^{\pm},\Pi,\bar \Pi]\,,\quad {\cal X}\in \{ \Lambda, \bar
\Lambda, \phi^\pm,\bar \phi^\pm\},
\end{eqnarray}
where each term in $\cal X$ contains at least one field from $
\{\varphi^\pm,\bar\varphi^{\pm},\Pi,\bar \Pi\}$ as in $X^{\rm New}$.
This is because all the new superpartners carry $R$ charges, and
$R$ charge should be conserved by the canonical transformation. In other words, $X^{\rm
New}[{\cal B}, \bar {\cal B}, \xi^\pm, \bar
\xi^\pm,\varphi^\pm,\bar\varphi^{\pm},\Pi,\bar \Pi]$ and ${\cal
X}[{\cal B}, \bar {\cal B}, \xi^\pm, \bar
\xi^\pm,\varphi^\pm,\bar\varphi^{\pm},\Pi,\bar \Pi]$   cannot
contribute to the QCD pieces containing only QCD fields in the
canonical transformation equation (\ref{eq:canonical-eq}). Therefore
$X^{\rm QCD}$ must separately satisfy (\ref{eq:canonical-eq-QCD}),
whilst the new pieces $X^{\rm New}$ along with $\cal X$ must satisfy
the canonical transformation equation with the pure QCD terms
eliminated:
\begin{eqnarray}
L^{+-}_A+ L^{+-}_{\Lambda}+ L^{+-}_{\phi}+ L^{+-}_{\alpha}
+ L^{++-}_{A}+ L^{++-}_{\Lambda A}&&
\nonumber\\
&&\hspace{-4cm}+ L^{++-}_{\alpha A}+ L^{++-}_{\phi
A}+ L^{++-}_{\alpha\Lambda \phi}
=
L^{+-}_{\varphi}+ L^{+-}_{\Pi}\,.
\label{eq:canonical-eq-new}
\end{eqnarray}
Here only terms containing the new pieces $X^{\rm New}$ in $L^{+-}_A,
L^{++-}_A,L^{+-}_{\alpha},L^{++-}_{\alpha A}$ are retained. So we
would expect $X^{\rm QCD}$  are just what we already have, which are listed in
appendix \ref{sect:sum-canonical-QCD}. 

To obtain the MHV lagrangian for massless pieces, we demand that all
the old plus chirality fields depend only on new plus chirality fields
and all the old minus chirality fields should depend linearly on the
new minus chirality fields. As in \cite{Ettle:2006bw,Ettle:2008ey},
first, we still demand that ${\cal A }$ is just a functional of ${\cal
B}$, {\it i.e.} in momentum space:\footnote{Further details on
definitions in the equations here and later, are included with the
summaries in appendices \ref{sect:sum-canonical-QCD} and
\ref{sect:sum-canonical-SQCD}.}
\begin{eqnarray}
{\cal A}_q &=& \sum_{n=1}^\infty\int_{1\cdots
n}\!\!\!\Upsilon_{q,\bar1\cdots\bar n}{\cal B}_{1}\cdots
{\cal B}_{n}\delta_{q\bar1\cdots\bar n}\,.
\label{eq:A}
\end{eqnarray}
It is then easy to obtain the canonical transformation for $\Lambda$ by
using the light-cone SUSY transformation (\ref{eq:lc-susy-6}) and
(\ref{eq:lc-susy-6-new}):
\begin{eqnarray}
{\Lambda}_q&=&\sum_{n=1}^\infty\int_{1\cdots
n}\!\!\!\Upsilon_{q,\bar1\cdots\bar n}\sum_{l=1}^{n}{\cal
B}_{1}\cdots\Pi_l\cdots
{\cal B}_{n}\delta_{q\bar1\cdots\bar n}\,.
\end{eqnarray}
Under SUSY, $\Lambda$ transforms back to ${\cal A}$ under the
light-cone SUSY transformation. Therefore the right hand side of the
above equation had better transform under SUSY back to the right hand
side of (\ref{eq:A}). After collecting terms, that is what happens,
thus forming a non-trivial consistency check on our reasoning.  As an
inverse, $\cal B$ is  only a functional of $\cal A$, {\it i.e.} ${\cal
B}[{\cal A}]$,  and $\Pi$ a functional of $\cal A$ and $\Lambda$, {\it
i.e.} $\Pi[{\cal A},\Lambda]$.

Next, let us consider  $\alpha^-$ and $\bar \phi^-$. The expansion of
$\alpha^-$ should at least contain one piece containing terms of the
form ${\cal B}\cdots{\cal B}\xi^-$. The supersymmetric transformation
of $\alpha^-$ and $\xi^-$ involves only terms proportional to
holomorphic $\eta$, but the supersymmetric transformation of ${\cal
B}$ involves only anti-holomorphic $\bar \eta$. To satisfy the SUSY
transformation for $\alpha^-$, the expansion must have another piece
to cancel the $\bar \eta$ terms after the SUSY transformation. But for
$\bar\phi^-$ there is no such requirement because the SUSY
transformation of $\bar \phi^-$ involves only $\bar \eta$ like ${\cal
B}$. So for a minimal extension of MHV-QCD,  the expansion of $\bar
\phi^-$ could contain only one piece:
\begin{eqnarray}
\bar\phi^-_q&=&\sum_{n=1}^\infty\int_{1\cdots
n}\!\!\!\Upsilon^-_{q,\bar1\cdots\bar n}{\cal B}_{1}\cdots
{\cal B}_{n-1}\bar\varphi_n^-\delta_{q\bar1\cdots\bar n}\,,
\label{eq:susy-phim}
\end{eqnarray}
and by using the SUSY transformation from $\phi^-$ to $\alpha^-$, the expansion
of $\alpha^-$ can be obtained:
\begin{eqnarray}
 \alpha^-_q&=&\xi^-_q+\sum_{n=2}^\infty\int_{1\cdots
n}\!\!\!\Upsilon^-_{q,\bar1\cdots\bar n}\Big( {\cal B}_{1}\cdots
{\cal B}_{n-1}\xi_n^--\sum_{l=1}^{n-1}{\cal
B}_{1}\cdots\Pi_l\cdots
{\cal B}_{n-1}\bar\varphi^-_n\Big)\delta_{q\bar1\cdots\bar n}\,.
\end{eqnarray}
Under SUSY, $\alpha^-$ transforms back to ${\bar\phi}^-$. After
collecting terms, one finds that the right hand side of the above
equation transforms under SUSY back to the right hand side of
(\ref{eq:susy-phim}). Again, this forms a non-trivial consistency check
on our reasoning.  As  a result, the inverse $\bar\varphi^-$ is only a
functional of $\bar\phi^-$ and ${\cal A}$, {\it i.e.}
$\bar\varphi^-[\bar\phi^-,{\cal A}]$, and $\xi^-$ a functional of
$\bar\phi^-$, $\Lambda$, $\alpha^-$, and $\cal A$, {\it i.e.}
$\xi^-[\alpha^-,\bar\phi^-,\Lambda,{\cal A}]$. The discussion for
$\bar\alpha^-$ and $\phi^-$ is the same.  For $\bar \alpha^+$,
$\bar\phi^+$ and $\alpha^+$, $\phi^+$, a similar discussion can be
applied, but the roles of $\alpha$ and $\phi$ are exchanged, that is,
the expansions for $\alpha^+$ and $\bar\alpha^+$ contain only one
piece  each and the expansions $\phi^+$ and $\bar \phi^+$ have two
pieces each. The detailed transformations are summarized in appendix
\ref{sect:sum-canonical-SQCD}. Here we only summarize the field
dependence:
\begin{eqnarray}
&&\alpha^+[\xi^+,{\cal B}]\,, \quad \bar\alpha^+[\bar\xi^+,{\cal
B}]\,,\quad \alpha^-[\xi^-,\bar\varphi^-,\Pi,{\cal B}]\,, \quad
\bar\alpha^-[\bar\xi^-,\varphi^-,\Pi,{\cal
B}]\,,
\nonumber\\  
&&
\phi^-[\varphi^-,{\cal B}]\,,\quad \bar\phi^-[\bar\varphi^-,{\cal
B}]\,,
\quad \phi^+[\varphi^+,\xi^+,\Pi,{\cal B}]\,,\quad
\bar\phi^+[\bar\varphi^+,\bar \xi^+,\Pi,{\cal
B}]\,,
\label{eq:field-dependence}
\end{eqnarray}
and the inverse field dependence:
\begin{eqnarray}
&&\xi^+[\alpha^+,{\cal A}]\,, \quad \bar\xi^+[\bar\alpha^+,{\cal
A}]\,,\quad \xi^-[\alpha^-,\bar\phi^-,\Lambda,{\cal A}]\,, \quad
\bar\xi^-[\bar\alpha^-,\phi^-,\Lambda,{\cal
A}]\,,
\nonumber \\  
&&
\varphi^-[\phi^-,{\cal A}]\,,\quad \bar\varphi^-[\bar\phi^-,{\cal
A}]\,,
\quad \varphi^+[\phi^+,\alpha^+,\Lambda,{\cal A}]\,,\quad
\bar\varphi^+[\bar\phi^+,\bar \alpha^+,\Lambda,{\cal
A}]\,.
\label{eq:inv-field-dependence}
\end{eqnarray}

The transformations for $\bar \Lambda $ and $\bar {\cal A}$ are  the
most complicated ones. From the transformation for MHV-QCD we can see
that, $\bar {\cal A}$ has at least two pieces: one piece involves only
gauge fields $\cal B$ and  $\bar {\cal B}$, and the other involves
fermions and their $1/N_C$ terms.  We look at the term involving $\cal
B$ and $\bar {\cal B}$ {\it i.e.} $\bar {\cal A}^{B}$ first. From a
similar holomorphic analysis of the SUSY transformation, one finds
that the minimal extension is to make the corresponding gauge piece
$\bar \Lambda^{\Pi B}$ of $\bar \Lambda$ contain a single piece
involving $\cal B$ and $\bar\Pi$, and $\bar{\cal A}$ contain an
additional piece ${\cal A}^{B\Pi}$ involving $\cal B$, $\Pi$, $\bar
\Pi$, in the following sense:
\begin{eqnarray}
\bar\Lambda^{B\Pi}_q&=&\sum_{n=1}^\infty\int_{1\cdots
n}\sum_{l=1}^{n}\Xi^{l}_{q,\bar1\cdots\bar n}{\cal
B}_{1}\cdots\bar\Pi_l\cdots
{\cal B}_{n}\delta_{q\bar1\cdots\bar n}
\,,\\
\bar{\cal A}^{B\Pi}_q&=&-\frac 1 {\sqrt2 \hat q}\sum_{n=2}^\infty\int_{1\cdots
n}\sum_{s=1}^{n}\bigg[\Xi^{s}_{q,\bar1\cdots\bar
n}\sum_{l=1,l\neq s}^{n}(-1)^{\delta_{ls}}{\cal
B}_{1}\cdots\Pi_l\cdots\bar \Pi_s\cdots
{\cal B}_{n}\bigg]\delta_{q\bar1\cdots\bar n}
\,.
\end{eqnarray}

Next, let us look at the piece involving ${\cal B}\cdots
\xi^+\bar\xi^-\cdots {\cal B}$ in the expansion of $\bar{\cal A}$.
There must be a corresponding piece in $\bar \Lambda$ which should be
transformed into this piece under the SUSY transformation. The
corresponding piece in $\bar\Lambda$ could contain terms proportional
to ${\cal B}\cdots\Pi\cdots \xi^+\bar \xi^-\cdots {\cal B}$, ${\cal
B}\cdots \varphi^+\bar\xi^-\cdots {\cal B}$ and ${\cal B}\cdots
\xi^+\varphi^-\cdots {\cal B}$. From (\ref{eq:inv-field-dependence})
and the requirement that this be a canonical transformation, we have: 
\begin{eqnarray}
\frac {\delta \bar \Lambda}{\delta\hat\partial\varphi^+}
=-\frac{\delta\varphi^-}{\delta\Lambda}=0\,,
\quad 
\frac {\delta \bar \Lambda}{\delta\bar\xi^-}
=\frac{\delta\xi^+}{\delta\Lambda}=0\,.
\end{eqnarray}
Therefore this piece cannot depend on $\varphi^+$ and $\bar \xi^-$,
and the only terms left are proportional to ${\cal B}\cdots
\xi^+\varphi^-\cdots {\cal B}$. This is also consistent with the SUSY
transformation since the SUSY transformations of ${\cal B}$,
$\varphi^-$ and $\xi^+$ involve only $\bar \eta$ which is the same
for $\bar \Lambda$.  By the same reasoning, one can obtain the
corresponding pieces in $\bar\Lambda$ for the other pieces in $\bar
{\cal A}$. In this way we demonstrate that  $\bar \Lambda$ must be
given by equations
(\ref{eq:canon-trans-barL-1})--(\ref{eq:canon-trans-barL-3}).  Now,
using SUSY transformation on $\bar \Lambda$ one determines the
canonical transformation for $\bar {\cal A}$ as given in the equations
(\ref{eq:canon-trans-barA-1})--(\ref{eq:canon-trans-barA-5}).
Finally, by a straightforward computation, one confirms that the
expansion for $\bar{\cal A}$ transforms back to the expansion for
$\bar\Lambda$ under the SUSY transformations.
 
We have now determined the canonical transformations for all of the
fields in terms of coefficients which we have tacitly assumed are the
same as the ones in the transformation to MHV-QCD. This assumption is
correct, since the QCD pieces of $\{\cal A, \bar {\cal A}, \alpha^\pm,
\bar \alpha^\pm\}$ must be just the same as the transformation for
MHV-QCD. It follows that the canonical transformations for new fields
does not introduce new unknown coefficients, as one might expect from
supersymmetry.

\section{The massive CSW vertices for SQCD}
\label{sect:CSW-massiveV-SQCD}
With the canonical transformations at hand, we are ready to look at
the new CSW vertices for SQCD. As proved in
\cite{Mansfield:2005yd,Ettle:2008ey}, the massless part of the CSW
vertices are the same as the MHV amplitude continued off shell up to
some external polarizations. These MHV vertices can be constructed
from the MHV amplitudes obtained using normal massless SUSY Ward
Identities (SWI), so we will not discuss them here. The new vertices
are the massive CSW vertices from mass terms in the light-cone SQCD.

The new massive CSW vertices for SQCD could be calculated by
substituting the canonical transformations into the light-cone SQCD
lagrangian. But we can try to fix them from the light-cone SUSY
transformation.  First let us look at 
\begin{eqnarray}
L^{+-}_{m,\alpha\phi}&=&L^{+-}_{m,\alpha}+L^{+-}_{m,\phi}.
\end{eqnarray}
Upon substitution of the canonical transformation, we see that this piece
should be composed of six kinds of field configurations:
\begin{eqnarray}
L^{+-}_{m,\alpha\phi}
&=&\frac{m^2}{\sqrt 2}\sum_{n=2}^{\infty}\int_{1\cdots
n}\bigg[
V^{+-}_{m,\xi}(\bar 1 \cdots \bar n)\bar\xi^+_{1}{\cal B}_{2}\cdots{\cal
B}_{n-1}\xi^-_n
+V^{-+}_{m,\xi}(\bar 1 \cdots \bar n)\,\bar\xi^-_{1}{\cal B}_{2}\cdots{\cal
B}_{n-1}\xi^+_n
\nonumber\\
&&+V^{+-}_{m,\varphi}(\bar1\cdots\bar n)\bar\varphi^+_{1}{\cal B}_{2}\cdots{\cal
B}_{n-1}\bar\varphi^-_n
+V^{-+}_{m,\varphi}(\bar1\cdots\bar n)\varphi^-_{1}{\cal B}_{2}\cdots{\cal
B}_{n-1}\varphi^+_n\bigg]\delta_{1\cdots n}
\nonumber\\
&&+\frac{m^2}{\sqrt 2}\sum_{n=3}^{\infty}\int_{1\cdots
n}\sum_{l=2}^{n-1}
\bigg[V^{l,+-}_{m,\xi\Pi\varphi}(\bar1\cdots\bar n)\,\bar\xi^+_{1}{\cal B}_{2}\cdots \Pi_l \cdots{\cal
B}_{n-1}\bar\varphi^-_n
\nonumber \\
&&+V^{l,-+}_{m,\varphi\Pi\xi}(\bar1\cdots\bar n)\,\varphi^-_{1}{\cal B}_{2}\cdots \Pi_l\cdots{\cal
B}_{n-1}\xi^+_n\bigg]\delta_{1\cdots n}\,.
\label{eq:Lm-pm}
\end{eqnarray}
Note that there are no $\bar\varphi^+{\cal B}\cdots \Pi\cdots {\cal
B}\xi^-$ and $\bar\xi^-{\cal B}\cdots \Pi\cdots {\cal B}\varphi^+$
terms. This is a consequence of  the simple field dependence of
$\alpha^+$, $\bar\alpha^+$, $\phi^-$, and $\bar\phi^-$ in
(\ref{eq:field-dependence}).  It also conforms to a rule that one can
extract from the light-cone SQCD lagrangian in appendix
\ref{sect:lc-SQCD-L}, namely that in the terms involving one scalar
and one gluino, the chiralities of the scalar and gluino are always
opposite.  The first, third and fifth pieces should be closed under
SUSY transformations. Using the SUSY transformations
(\ref{eq:lc-susy-1-new})--(\ref{eq:lc-susy-6-new}) one then easily
finds that these three vertices must be related as:
\begin{eqnarray}
V^{+-}_{m,\varphi} =-\sqrt 2\, \hat 1 V^{+-}_{m,\xi} 
\,,\quad
V^{l,+-}_{m,\xi\Pi\varphi}= V^{+-}_{m,\xi}\,. 
\label{eq:V+-m}
\end{eqnarray}
Since only the QCD pieces in $\alpha^\pm$ and $\bar \alpha^\pm$ can
contribute to terms proportional to $\bar\xi^+_{1}{\cal B}_{2}\cdots{\cal
B}_{n-1}\xi^-_n$ and $\bar\xi^-_{1}{\cal B}_{2}\cdots{\cal
B}_{n-1}\xi^+_n$, $V^{+-}_{m,\xi}$ and $V^{-+}_{m,\xi}$  must be equal to the
corresponding coefficients for MHV-QCD.  
By using the results of $V^{+-}_{m,\xi}$ for MHV-QCD listed in
the appendix \ref{sect:mass-V-QCD}, the above relations immediately
give $V^{+-}_{m,\varphi}$ and $V^{+-}_{m,\xi\Pi\varphi}$. 
Similarly, we can also obtain expressions for the other
three vertices:
\begin{eqnarray}
V^{-+}_{m,\varphi} =\sqrt 2\, \hat 1 V^{-+}_{m,\xi} 
\,,\quad
V^{l,-+}_{m,\varphi\Pi\xi}= V^{-+}_{m,\xi}\,. 
\label{eq:V-+m}
\end{eqnarray}
If one calculates the amplitudes $A(1_{\rm q}^\pm2^+\cdots (n-1)^+
n_{\bar {\rm q}}^\mp)$ and $A(1_\varphi^\pm2^+\cdots (n-1)^+ n_\varphi
^\mp)$ by choosing the reference momenta of fermions as in
\cite{Schwinn:2008fm}, the amplitudes are proportional to
$V^{\pm\mp}_{m,\xi}$ and $V^{\pm\mp}_{m,\varphi}$. Using relation
(\ref{eq:V+-m}) and (\ref{eq:V-+m}), one can recover the relations
between these amplitudes obtained from massive SWI in
\cite{Schwinn:2008fm}. Notice that $V^{l,-+}_{m,\varphi\Pi\xi}$ and
$V^{l,+-}_{m,\xi\Pi\varphi}$ are independent of $l$.  This is a
consequence of the supersymmetry. Taking $V^{l,+-}_{m,\xi\Pi\varphi}$
as an example, since after SUSY transformation,  there is a  term
proportional to $ \bar\xi^+_1{\cal B}_2\cdots\Pi_s\cdots
\Pi_l\cdots{\cal B}\bar\varphi^-_n$ which comes  from
$V^{l,+-}_{m,\xi\Pi\varphi} \bar\xi^+_1{\cal B}_2\cdots \Pi_l\cdots{\cal
B}\bar\varphi^-_n$ and $V^{s,+-}_{m,\xi\Pi\varphi} \bar\xi^+_1{\cal
B}_2\cdots \Pi_s\cdots{\cal B}\bar\varphi^-_n$, whose coefficient must
vanish, we have:
\begin{equation}
V^{l,+-}_{m,\xi\Pi\varphi}-V^{s,+-}_{m,\xi\Pi\varphi}=0\,, \quad
\text{for } s\neq l\,,
\end{equation}
in other words $V^{l,+-}_{m,\xi\Pi\varphi}$ does not depend on
$l$.

Next, let us look at 
\begin{eqnarray}
L_m^{+-+}=L_{m,\alpha A}^{+-+}+L_{m,\phi A}^{+-+}+L_{m,\phi
\Lambda\alpha}^{+-+}\,.
\end{eqnarray}
We can separate it into three parts which are invariant under the supersymmetry:
\begin{eqnarray}
L_m^{+-+,(1)}&=& im\bigg\{\sum_{n=3}^{\infty}\int_{1\cdots
n}\sum_{s=2}^{n-1}\Big(V^{s,+-+}_{m,\xi}(\bar1\cdots\bar n)\,\bar\xi^+_{1}{\cal
B}_{2} \cdots \bar{\cal B}_{s}\cdots{\cal
B}_{n-1}\xi^+_n
\nonumber \\
&&+V^{s,+-+}_{m,\varphi\Pi\xi}(\bar1\cdots\bar n) \bar\varphi^+_{1}{\cal B}_{2} \cdots \bar\Pi_{s}\cdots{\cal
B}_{n-1}\xi^+_n
\nonumber \\
&&+V^{s,+-+}_{m,\xi\Pi\varphi}(\bar1\cdots\bar n)\, \bar\xi^+_{1}{\cal B}_{2} \cdots \bar\Pi_{s}\cdots{\cal
B}_{n-1}\varphi^+_n\Big)\delta_{1\cdots n}
\nonumber \\
&&+\frac1{\sqrt 2}\sum_{n=4}^{\infty}\int_{1\cdots
n}
\sum_{l=2}^{n-1}\sum_{s=2,s\neq l}^{n-1}(-1)^{\delta_{ls}}\Big(V^{ls,+-+}_{\xi\Pi\bar\Pi\xi}(\bar1\cdots\bar n)
\nonumber\\
&&\quad\times\bar\xi^+_{1}{\cal B}_{2} \cdots \Pi_{l}\cdots
\bar\Pi_{s}\cdots{\cal B}_{n-1}\xi^+_n\Big)\delta_{1\cdots n}
\bigg\}\,,
\label{eq:Lm-pmp-1}
\end{eqnarray}
\begin{eqnarray}
L_m^{+-+,(2)}&=& i\frac {mg^2}{2\sqrt 2}\bigg\{\sum_{n=4}^{\infty}\int_{1\cdots
n}\sum_{s=2}^{n-2}
\Big(V^{s,++-+}_{m,\xi}(\bar1\cdots \bar n)
\bar\xi^+_{1}{\cal
B}_{2} \cdots {\xi}^+_{s}\bar\xi^-_{s+1}\cdots{\cal
B}_{n-1}\xi^+_n
\nonumber \\
&&+V^{s,++-+}_{m,\xi\varphi\varphi\xi}(\bar1\cdots \bar n)\,\bar\xi^+_{1}{\cal
B}_{2} \cdots {\varphi}^+_{s}\varphi^-_{s+1}\cdots{\cal
B}_{n-1}\xi^+_n
\nonumber \\
&&+V^{s,++-+}_{m,\varphi\xi\varphi\xi}(\bar1\cdots \bar n)\, \bar\varphi^+_{1}{\cal B}_{2}
\cdots\xi^+_s \varphi^-_{s+1}\cdots{\cal
B}_{n-1}\xi^+_n
\nonumber \\
&&+V^{s,++-+}_{m,\xi\xi\varphi\varphi}(\bar1\cdots \bar n)\bar\xi^+_{1}{\cal B}_{2}
\cdots\xi^+_s \varphi^-_{s+1}\cdots{\cal
B}_{n-1}\varphi^+_n\Big)\delta_{1\cdots n}
\nonumber \\
&&+\sum_{n=5}^{\infty}\int_{1\cdots
n}\sum_{s=2}^{n-2}\sum_{l=2,l\neq
s,s+1}^{n-1}\!\!\!(-1)^{\delta_{ls}}\Big(V^{ls,++-+}_{m,\xi\Pi\varphi\varphi\xi}(\bar1\cdots
\bar n)
\nonumber \\
&&\quad \times\bar\xi^+_{1}{\cal B}_{2} \cdots \Pi_{l}\cdots
\xi^+_{s}\varphi^-_{s+1}\cdots{\cal B}_{n-1}\xi^+_n\Big)\delta_{1\cdots n}
\bigg\}\,,
\label{eq:Lm-pmp-2}
\end{eqnarray}
\begin{eqnarray}
L_m^{+-+,(3)}&=& i\frac {mg^2}{2\sqrt 2}\bigg\{\sum_{n=4}^{\infty}\int_{1\cdots
n}\sum_{s=2}^{n-2}
\Big(V^{s,+-++}_{m,\xi}(\bar1\cdots \bar n)\bar\xi^+_{1}{\cal
B}_{2} \cdots {\xi}^-_{s}\bar\xi^+_{s+1}\cdots{\cal
B}_{n-1}\xi^+_n
\nonumber \\
&&+ V^{s,+-++}_{m,\xi\varphi\varphi\xi}(\bar1\cdots \bar n)\,\bar\xi^+_{1}{\cal
B}_{2} \cdots \bar{\varphi}^-_{s}\bar \varphi^+_{s+1}\cdots{\cal
B}_{n-1}\xi^+_n
\nonumber \\
&&+V^{s,+-++}_{m,\varphi\varphi\xi\xi}(\bar1\cdots \bar n)\bar\varphi^+_{1}{\cal B}_{2}
\cdots\bar\varphi^-_s \bar\xi^+_{s+1}\cdots{\cal
B}_{n-1}\xi^+_n
\nonumber \\
&&+V^{s,+-++}_{m,\xi\varphi\xi\varphi}(\bar1\cdots \bar n)\bar\xi^+_{1}{\cal B}_{2}
\cdots\bar\varphi^-_s \bar\xi^+_{s+1}\cdots{\cal
B}_{n-1}\varphi^+_n\Big)\delta_{1\cdots n}
\nonumber \\
&&+\sum_{n=5}^{\infty}\int_{1\cdots
n}
\sum_{s=2}^{n-2}\sum_{l=2,l\neq
s,s+1}^{n-1}(-1)^{\delta_{ls}}\Big(V^{ls,+-++}_{m,\xi\Pi\varphi\xi\xi}(\bar1\cdots
\bar n)
\nonumber \\
&&\quad\times \bar\xi^+_{1}{\cal B}_{2} \cdots \Pi_{l}\cdots
\bar\varphi^-_{s}\bar\xi^+_{s+1}\cdots{\cal B}_{n-1}\xi^+_n\Big)\delta_{1\cdots n}
\bigg\}\,,
\label{eq:Lm-pmp-3}
\end{eqnarray}
where $\delta_{ls}$ is defined in (\ref{eq:delta-ls}). Just by using
the SUSY transformation as before, we arrive at the following
relations between these vertices:
\begin{eqnarray}
&&V^{s,+-+}_{m,\varphi\Pi\xi}=\frac{\hat1}{\hat s}V^{s,+-+}_{m,\xi}\,,\quad
V^{s,+-+}_{m,\xi\Pi\varphi}=-\frac{\hat n}{\hat s}V^{s,+-+}_{m,\xi}\,,\quad
V^{ls,+-+}_{m,\xi\Pi\bar\Pi\xi}=-\frac{1}{\hat s}V^{s,+-+}_{m,\xi}
\nonumber\\
&&V^{s,++-+}_{m,\varphi\xi\varphi\xi}=\sqrt2 \hat 1 V^{s,++-+}_{m,\xi}\,,\quad
V^{s,++-+}_{m,\xi\xi\varphi\varphi}=-\sqrt2 \hat n V^{s,++-+}_{m,\xi}\,,\quad
\nonumber \\
&&V^{s,++-+}_{m,\xi\varphi\varphi\xi}=\sqrt2 \hat s V^{s,++-+}_{m,\xi}\,,\quad
V^{ls,++-+}_{m,\xi\Pi\varphi\varphi\xi}=-V^{s,++-+}_{m,\xi}\,,
\nonumber\\
&&V^{s,+-++}_{m,\xi\varphi\xi\varphi}=-\sqrt2 \hat n V^{s,+-++}_{m,\xi}\,,\quad
V^{s,+-++}_{m,\varphi\varphi\xi\xi}=\sqrt2 \hat 1 V^{s,+-++}_{m,\xi}\,,\quad
\nonumber \\
&&V^{s,+-++}_{m,\xi\varphi\varphi\xi}=-\sqrt2 \widehat{s+1} V^{s,+-++}_{m,\xi}\,,\quad
V^{ls,+-++}_{m,\xi\Pi\varphi\varphi\xi}=-V^{s,+-++}_{m,\xi}\,.
\end{eqnarray}
Since all the vertices on the right hand side of these equations have
already been obtained from MHV-QCD, the new vertices on the left hand
side immediately follow from these relations. Notice that
$V^{ls,+-+}_{m,\xi\Pi\bar\Pi\xi}$, 
$V^{ls,++-+}_{m,\xi\Pi\varphi\varphi\xi}$ and
$V^{ls,+-++}_{m,\xi\Pi\varphi\varphi\xi}$ also do not depend on $l$,
so the dependence of $l$ in corresponding terms  appears only in
$(-1)^{\delta_{ls}}$, {\it i.e.} when the positions of plus-helicity
gluinos change, the vertices at most change sign, in accordance with
Fermi statistics. Just like in the previous case, this can be
understood as a result of the supersymmetry: for example, the
coefficient of the term $\bar \xi ^+\cdots \Pi_{l_1}\cdots
\Pi_{l}\cdots \xi^+\varphi^-\cdots\xi^+$, generated by the SUSY
transformation, must vanish.  This arises from transforming the
$V^{ls,++-+}_{m,\xi\Pi\varphi\varphi\xi}$,
$V^{l_1s,++-+}_{m,\xi\Pi\varphi\varphi\xi}$ terms, from which we
conclude
\begin{eqnarray}
V^{ls,++-+}_{m,\xi\Pi\varphi\varphi\xi}-V^{l_1s,++-+}_{m,\xi\Pi\varphi\varphi\xi}=0
\,, \quad \text{for } l_1\neq l\,,
\end{eqnarray}
which just means that $V^{ls,++-+}_{m,\xi\Pi\varphi\varphi \xi}$ is
independent of $l$.

Now let us look at the last piece
\begin{eqnarray}
L_{m}^{-+-}&=&L_{m,\alpha A}^{-+-}+L_{m,\phi\Lambda\alpha}^{-+-}\,. 
\end{eqnarray}
There are four kinds of vertices with different field configurations in the
lagrangian after applying the canonical transformations: 
\begin{eqnarray}
L_{m}^{-+-}&=&i m\bigg \{ \sum_{n=3}^{\infty}\int_{1\cdots
n}\bigg[V^{-+-}_{m,\xi}(\bar1\cdots\bar n)\bar\xi^-_{1}{\cal B}_{2} \cdots{\cal
B}_{n-1}\xi^-_n
\nonumber \\
&&+\sum_{l=2}^{n-1}V^{l,-+-}_{m,\varphi\Pi\xi}(\bar1\cdots\bar n)\varphi^-_{1}{\cal B}_{2} \cdots \Pi_l\cdots{\cal
B}_{n-1}\xi^-_n
\nonumber \\
&&+\sum_{l=2}^{n-1}V^{l,-+-}_{m,\xi\Pi\varphi}(\bar1\cdots\bar n)\bar\xi^-_{1}{\cal B}_{2} \cdots \Pi_l\cdots{\cal
B}_{n-1}\bar\varphi^-_n
\bigg]\delta_{1\cdots n}
\nonumber \\
&&\hspace{-1cm}+\sum_{n=4}^{\infty}\int_{1\cdots
n}
\sum_{l_1=2}^{n-2}\sum_{l_2=l_1+1}^{n-1}V^{l_1l_2,-+-}_{m,\varphi\Pi\Pi\varphi}(\bar1\cdots\bar n)
\varphi^-_{1}{\cal B}_{2} \cdots \Pi_{l_1}\cdots \Pi_{l_2}\cdots{\cal
B}_{n-1}\bar\varphi^-_n\:\delta_{1\cdots n}
\bigg\}
\,.\end{eqnarray}
Unfortunately, the light-cone SUSY transformation can only provide some
relations among these vertices:
\begin{eqnarray}
&&\hat 1 \,V^{-+-}_{m,\xi}+\sum_{l=2}^{n-1}\hat l\,
V^{l,-+-}_{m,\varphi\Pi\xi}=0
\,,\quad\hat n \,V^{-+-}_{m,\xi}-\sum_{l=2}^{n-1}\hat l\,
V^{l,-+-}_{m,\xi\Pi\varphi}=0
\,,
\label{eq:V-+--relations-1}\\
&&V^{-+-}_{m,\xi}-V^{l,-+-}_{m,\varphi\Pi\xi}+ V^{l,-+-}_{m,\xi\Pi\varphi}=0
\,,\label{eq:V-+--relations-2}\\
&&\hat n\,V^{l,-+-}_{m,\varphi\Pi\xi}+\hat
1\,V^{l,-+-}_{m,\xi\Pi\varphi}+ \sum_{l_1=2}^{l-1}\hat
l_1V^{l_1l,-+-}_{m,\varphi\Pi\Pi\varphi}-\sum_{l_1=l+1}^{n-1}\hat
l_1V^{ll_1,-+-}_{m,\varphi\Pi\Pi\varphi}=0
\,,
\label{eq:V-+--relations-3}\\&&
V^{l_2,-+-}_{m,\varphi\Pi\xi}- V^{l_1,-+-}_{m,\varphi\Pi\xi}-V^{l_1l_2,-+-}_{m,\varphi\Pi\Pi\varphi}=0
\,,\label{eq:V-+--relations-4}\\&&
V^{l_2,-+-}_{m,\xi\Pi\varphi}- V^{l_1,-+-}_{m,\xi\Pi\varphi}-V^{l_1l_2,-+-}_{m,\varphi\Pi\Pi\varphi}=0
\,,\label{eq:V-+--relations-5}\\
&&
V^{l_1l_2,-+-}_{m,\varphi\Pi\Pi\varphi}-V^{sl_2,-+-}_{m,\varphi\Pi\Pi\varphi}+V^{sl_1,-+-}_{m,\varphi\Pi\Pi\varphi}=0\,.
\label{eq:V-+--relations-6}
\end{eqnarray}
From these relations, we cannot fix these vertices uniquely, but the
above relations may simplify their determination. Indeed, we only need
to calculate two of these vertices to obtain the others. Since we
already have $V^{-+-}_{m,\xi}$ for MHV-QCD, actually we only need to
calculate one vertex. The results are:
\begin{eqnarray}
V^{l,-+-}_{m,\varphi\Pi\xi}&=&\frac{\hat 1\:(l\:n)}{\hat l\:(1\:n)}V^{-+-}_{m,\xi}\,,
\label{eq:V-+-1}\\
V^{l,-+-}_{m,\xi\Pi\varphi}&=&-\frac{\hat n\:(1\:l)}{\hat l\:(1\:n)}V^{-+-}_{m,\xi}
\,,
\label{eq:V-+-2}\\
V^{l_1l_2,-+-}_{m,\varphi\Pi\Pi\varphi}&=&-\frac{\hat1\hat n\:(l_1\:l_2)}{\hat l_1\hat l_2\:(1\:n)}V^{-+-}_{m,\xi}
\,.\label{eq:V-+-3}
\end{eqnarray}
It is easy to check that these equations satisfy
(\ref{eq:V-+--relations-1})--(\ref{eq:V-+--relations-6}). These
relations in fact can also be understood using the massive SWI at the
amplitude level \cite{Schwinn:2008fm}.

\section{Conclusion and discussion}
\label{sect:conclusion}
In this paper, we have seen that the whole canonical transformation
for the MHV-SQCD lagrangian can be obtained simply by applying a kind
of light-cone supersymmetry transformation on the canonical
transformation for MHV-QCD lagrangian. Unlike SWI, this SUSY
transformation relates the dynamical fields of light-cone SQCD
directly at the lagrangian level, not the annihilation operators of
outgoing states at the amplitude level. As a result, it is more widely
applicable, as we saw for example by using it to uniquely determine
the canonical transformation for MHV-SQCD. Some relations among
massive CSW vertices can be understood using this supersymmetric
transformation. Using these relations and the canonical
transformation, all the massive CSW vertices are obtained. But since
this SUSY transformation is a subgroup of the whole supersymmetry
transformation, it is not as flexible as SWI in changing the
transformation parameters. We see this in the massive vertex relations
(\ref{eq:V-+-1})--(\ref{eq:V-+-3}), which can be obtained from SWI at
the amplitude level, but can not all be obtained just by using this
SUSY transformation. 

Though in this paper, the canonical transformation is derived for
MHV-SQCD with one flavour, clearly its application is more general.
More flavours can be added without difficulty.  Since the pieces
having different field content in the canonical transformation
equation (\ref{eq:canonical-eq}) cancel separately from the left and
right hand side of the equation, parts of the transformation can be
directly used in theories which can be embedded in SQCD. A typical
example is MHV-QCD. After  turning off $\Lambda$ and $\phi_{1,2}$ by
setting them to zero,  we can obtain the canonical transformation for
MHV-QCD and the corresponding massive CSW vertices are not changed.
For theories involving only a gauge field and scalars, we can set
$\Lambda$ and quark fields to be zero in the canonical transformation,
and the corresponding parts of the massive CSW vertices can be used
directly in this theory as in \cite{Boels:2007pj,Boels:2008ef}. This
provides a direct explanation of the similarity between the massive
CSW scalar vertices in \cite{Boels:2007pj,Boels:2008ef} and the
massive CSW fermion vertices in \cite{Ettle:2008ey}. Moreover,
changing the masses for scalars and fermions does not modify the
canonical transformation but the CSW vertices could be modified. It is
also possible to extend it to a supersymmetric theory incorporating
standard model such as the MSSM.

These CSW vertices  of course can be used in simplifying calculations
of SQCD amplitudes involving massive quarks and scalars.  An
interesting observation is that for terms involving a plus-helicity
gluino and containing only one minus-chirality particle, the massive
vertices do not depend on the position of the plus-helicity gluino in
the color matrix product except for a possible sign change according
to fermion statistics. This is a consequence of supersymmetry. This
property may be useful in the amplitude calculations in SQCD.

\appendix

\section*{Acknowledgments}
The authors  thank the STFC for financial support.

\section{Light-cone lagrangian for SQCD}
\label{sect:lc-SQCD-L}
\begin{eqnarray}
L_{A}^{+-} &=& \frac 4{g^2}\,{\rm tr}\!\int_\Sigma\!\! d^3{\bf x}\,\,
{\bar {\cal A}}\,(\check\partial\hat\partial-
\partial\bar\partial)\,{\cal A}
\,,\\
L^{+-}_\Lambda&=&-{i}\frac{2\sqrt2}{g^2}{\rm tr}\!\int_\Sigma\!\!
d^3{\bf x}\,\,(\bar\Lambda\check{\partial}\Lambda
-\bar\Lambda\bar\partial\hat\partial^{-1}{\partial}\Lambda
)
\,,\\
L^{+-}_\phi&=&-2\!\int_\Sigma\!\! d^3{\bf x}\,\,(\bar \phi^+(\hat \partial \check\partial -\partial
\bar \partial)\bar \phi^-+\phi^-(\hat \partial \check\partial -\partial
\bar \partial)\phi^+)
\,,\\
L^{+-}_\alpha&=&{i}{\sqrt2}\!\int_\Sigma\!\! d^3{\bf x}\,\,(\bar\alpha^+\check{\partial}\alpha^-+\bar\alpha^-\check{\partial}\alpha^+
-\bar\alpha^+\bar\partial\hat\partial^{-1}{\partial}\alpha^-
-\bar\alpha^-\partial\hat\partial^{-1}\bar{\partial}\alpha^+)
\,.
\end{eqnarray}

\begin{eqnarray}
{L}_{A}^{++-}&=&-\frac 4 {g^2}\,{\rm tr}\!\int_\Sigma\!\! d^3{\bf x}\,\,
({\bar\partial}{\hat\partial}^{-1} {\cal A})\,
[{\cal A},\,{\hat\partial} {\bar {\cal A}}]
\,,\\
{L}_{\Lambda A}^{++-}&=&-i\frac {2\sqrt2} {g^2}\,{\rm tr}\!\int_\Sigma\!\! d^3{\bf x}\,\,
\Big({\cal A}\{\Lambda,{\bar\partial}{\hat\partial}^{-1}
{\Lambda}\}-({\bar\partial}{\hat\partial}^{-1} {\cal A})\{\Lambda,\bar\Lambda\}\Big)\,
\,,\\
L_{\alpha A}^{++-}&=& {i\sqrt
2}\int_\Sigma\!\! d^3{\bf x}\,\,\Big(\bar\alpha^+(\hat\partial^{-1}\bar\partial {\cal A})\alpha^-
+\bar\alpha^-(\hat\partial^{-1}\bar\partial {\cal A})\alpha^+
\nonumber\\&&
-\bar\alpha^+\bar\partial \hat\partial^{-1}({\cal A}\alpha^-)
-\bar\alpha^-{\cal A}\hat\partial^{-1}\bar\partial \alpha^+\Big) 
\,,\\
L_{\phi A}^{++-}&=&-{ 
2}\int_\Sigma\!\! d^3{\bf
x}\,\,\Big(\bar\phi^+(\hat\partial^{-1}\bar\partial {\cal A})\hat
\partial \bar\phi^-
-\partial \phi^-(\hat\partial^{-1}\bar\partial {\cal A})\phi^+
\nonumber\\&&
-\bar\phi^+{\cal A}\bar \partial \bar \phi^-
+\bar\partial\phi^-{\cal A} \phi^+\Big) 
\,,\\
L_{\alpha\Lambda \phi }^{++-}&=&{i\sqrt 
2}\int_\Sigma\!\! d^3{\bf
x}\,\,\Big(-\bar\alpha^+(\hat\partial^{-1}\bar\partial {\Lambda})
\bar \phi^-
+ \phi^-(\hat\partial^{-1}\bar\partial {\Lambda})\alpha^+
\nonumber\\&&
+(\hat\partial^{-1}\bar\partial \bar\alpha^+){\Lambda}  \bar \phi^-
-\phi^-{\Lambda} (\hat\partial^{-1}\bar\partial\alpha^+) \Big) 
\,.
\end{eqnarray}

\begin{eqnarray}
{L}_{A}^{--+}&=&-\frac 4 {g^2} \,{\rm tr}\!\int_\Sigma\!\! d^3{\bf x}\,\, [{\bar{
\cal A}},\,{\hat\partial} {\cal  A}]\,
({  \partial}{\hat\partial}^{-1} {\bar {\cal A}})
\,,\\
{L}_{\Lambda A}^{--+}&=&-i\frac {2\sqrt2} {g^2} \,{\rm tr}\!\int_\Sigma\!\! d^3{\bf x}\,\, \bigg({\bar
{\cal A}}\,\{\Lambda,{  \partial}{\hat\partial}^{-1} {\bar \Lambda}\} -({  \partial}{\hat\partial}^{-1} {\bar {\cal A}})\{\bar \Lambda,\Lambda\}\bigg)
\,,\\
L_{\alpha A}^{--+}&=&i\sqrt2 \,\int_\Sigma\!\! d^3{\bf
x}\,\,\Big(\bar\alpha^+(\hat\partial^{-1}\partial \bar {\cal A})\alpha^-
+\bar\alpha^-(\hat\partial^{-1}\partial\bar {\cal A})\alpha^+
\nonumber\\&&
-\bar\alpha^-\partial \hat\partial^{-1}(\bar {\cal A}\alpha^+)
-\bar\alpha^+\bar {\cal A}(\hat\partial^{-1}\partial \alpha^-)\Big) 
\,,\\
L_{\phi A}^{--+}&=&2 \,\int_\Sigma\!\! d^3{\bf x}\,\,\Big(\hat
\partial\bar\phi^+(\hat\partial^{-1}\partial \bar {\cal A})\bar\phi^-
-\phi^-(\hat\partial^{-1}\partial\bar {\cal A})\hat\partial\phi^+
-\phi^-\bar {\cal A}\partial\phi^+
+\partial\bar\phi^+\bar {\cal A}\bar \phi^-\Big) 
\,,\\
L_{\alpha\Lambda \phi}^{--+}&=&-i\sqrt2 \,\int_\Sigma\!\! d^3{\bf x}\,\,\Big(\bar\phi^+(\hat\partial^{-1}\partial \bar \Lambda)\alpha^-
-\bar\alpha^-(\hat\partial^{-1}\partial\bar \Lambda)\phi^+
\nonumber\\&&
-\bar\phi^+\bar \Lambda(\hat\partial^{-1}\partial \alpha^-)
+(\hat\partial^{-1}\partial\bar\alpha^-)\bar \Lambda \phi^+\Big) 
\,.
\end{eqnarray}

\begin{eqnarray}
L^{--++}_{A}&=& \int_\Sigma\!\! d^3{\bf x}\,\,\Big[\frac 1
2\,\Sigma^a_A\hat\partial^{-2}\Sigma_A^a-\frac 1 {g^2}{\rm
tr}\big([{\cal A},\bar {\cal A}]^2\big)\Big ]
\,,\\
L^{--++}_{\Lambda}&=&\frac 1 2\,\int_\Sigma\!\! d^3{\bf
x}\,\,\Sigma^a_\Lambda\hat\partial^{-2}\Sigma_\Lambda^a
\,,\\
L^{--++}_{\Lambda A}&=&\int_\Sigma\!\! d^3{\bf
x}\,\,\bigg[\Sigma^a_\Lambda\hat\partial^{-2}\Sigma_A^a
-i\frac {2\sqrt 2} {g^2}{\rm tr}
\big([{\cal A},\bar {\Lambda}]\hat \partial^{-1}[\bar{\cal
A},{\Lambda}]\big)\bigg ]
\,.
\end{eqnarray}
\begin{eqnarray}
L^{--++}_{\phi}&=&\frac 12\int_\Sigma\!\! d^3{\bf x}\,\,\bigg[
\Sigma_\phi^a\hat \partial^{-2} \Sigma_\phi^a +g^2\,(-\bar \phi^+
T^a\bar \phi^- +\phi^- T^a \phi^+)^2\bigg]
\,,\\
L^{--++}_{\alpha}&=&\frac 1 2\,\int_\Sigma\!\! d^3{\bf
x}\,\,\Sigma^a_\alpha\hat\partial^{-2}\Sigma_\alpha^a
\,,\\
L^{--++}_{\alpha \phi}&=&\int_\Sigma\!\! d^3{\bf
x}\,\,\bigg[ \Sigma^a_\alpha\hat\partial^{-2}\Sigma_\phi^a 
\nonumber \\
&&-\sqrt 2 i
g^2(\bar \alpha^+
T^a\bar \phi^- +\phi^- T^a \alpha^+)\hat \partial^{-1}(\bar \phi^+
T^a \alpha^- +\bar\alpha^- T^a \phi^+)\bigg]
\,.\end{eqnarray}
\begin{eqnarray}
L^{--++}_{\phi A}&=&\int_\Sigma\!\! d^3{\bf x}\,\,\bigg[
\Sigma_\phi^a\hat \partial^{-2} \Sigma_A^a +\big(\bar \phi^+
\{ {\cal A},\bar{\cal A}\} \bar \phi ^{-} +\phi^- \{ {\cal
A},\bar{\cal A}\} \phi^+\big)\bigg]
\,,\\
L^{--++}_{\alpha\Lambda}&=&\int_\Sigma\!\! d^3{\bf
x}\,\,\Sigma^a_\alpha\hat\partial^{-2}\Sigma_\Lambda^a
\,,\\
L^{--++}_{\alpha A}&=&\int_\Sigma\!\! d^3{\bf
x}\,\,\bigg[\Sigma^a_\alpha\hat\partial^{-2}\Sigma_A^a
-i\sqrt 2\Big(\bar \alpha^-{\cal A}\hat\partial^{-1}(\bar {\cal
A}\alpha^+)+\bar \alpha^+\bar {\cal A}\hat \partial^{-1}({\cal A}
\alpha^-)\Big)\bigg ]
\,,\\
L^{--++}_{\phi \Lambda}&=&\int_\Sigma\!\! d^3{\bf
x}\,\,\bigg[\Sigma^a_\phi\hat\partial^{-2}\Sigma_\Lambda^a
-i\sqrt 2\Big(\phi^- \Lambda\hat\partial^{-1}(
\bar\Lambda\phi^+)+\bar \phi^+\bar\Lambda \hat \partial^{-1}(\Lambda
\bar\phi^-)\Big)\bigg ]
\,,\\
L^{--++}_{\alpha\phi \Lambda A}&=&-i\sqrt 2\int_\Sigma\!\! d^3{\bf
x}\,\,\bigg[
\phi^- \Lambda\hat\partial^{-1}(
\bar{\cal A}\alpha^+)-\bar \phi^+\bar\Lambda \hat \partial^{-1}({\cal A} \alpha^-)
-\hat \partial^{-1}(\bar\alpha^-{\cal A})\bar
\Lambda\phi^+
\nonumber\\
&&
+\hat \partial^{-1}(\bar\alpha^+\bar{\cal A}) \Lambda\phi^- 
+\bar \alpha^+
\big(\hat \partial^{-1}[ \bar{\cal A},\Lambda]\big) \bar \phi ^{-}
-\phi^- \big( \hat \partial^{-1}[ \bar{\cal A},\Lambda]\big)
\alpha^+
\nonumber \\
&&+\bar\phi^+ \big( \hat \partial^{-1}[ {\cal A},\bar\Lambda]\big) \alpha^-
-\bar \alpha^-
\big(\hat \partial^{-1}[ {\cal A},\bar\Lambda]\big) \phi ^{+}
\bigg]
\,.
\end{eqnarray}
where 
\begin{eqnarray}
\Sigma_A^a&=&-\frac2 g{\rm tr}\Big([{\cal A},\hat \partial \bar{\cal
A}]T^a+[\bar {\cal A},\hat \partial {\cal A}]T^a\Big)\,,
\\
\Sigma_\Lambda^a&=&\frac{i 2\sqrt2} g{\rm tr}\Big(\{\bar \Lambda,\Lambda\}
T^a\Big)
\\
\Sigma_\alpha^a&=&i\sqrt2 g\Big ( \bar \alpha^- T^a \alpha^++\bar
\alpha^+T^a \alpha^-\Big)
\\
\Sigma_\phi^a&=&-g\Big(-\hat\partial \bar \phi^+ T^a \bar\phi^-
+\bar \phi^+ T^a \hat \partial\bar\phi^-
+ \phi^- T^a \hat \partial\phi^+
-\hat \partial\phi^- T^a \phi^+\Big)
\end{eqnarray}
\begin{eqnarray}
L_{m,\alpha}^{+-}&=&i\frac {m^2}{\sqrt2}\int_\Sigma\!\! d^3{\bf
x}\,\,(\bar \alpha^-\hat
\partial^{-1}\alpha^++\bar\alpha^+\hat\partial^{-1} \alpha^-)
\,,
\label{eq:lcqcd-quarkmass} \\
L_{m,\phi}^{+-}&=&- {m^2}\int_\Sigma\!\! d^3{\bf
x}\,\,( \bar\phi^+
\bar\phi^-+\phi^- \phi^+)
\,,
\\
L_{m,\alpha A}^{+-+}&=&-m\int_\Sigma\!\! d^3{\bf
x}\,\,\bar\alpha^+[\hat\partial^{-1},\bar {\cal A}] \alpha^+
\,, 
\label{eq:lcqcd-massint1}
\\
L_{m,\phi\Lambda\alpha}^{+-+}&=&m\int_\Sigma\!\! d^3{\bf
x}\,\,\Big((\hat\partial^{-1}\bar\alpha^+)\bar
\Lambda \phi^+ -\bar \phi^+\bar \Lambda\hat \partial^{-1} \alpha^+\Big)\,, 
\label{eq:lcsqcd-massint1}
\\
L_{m,\alpha
A}^{-+-}&=&m\int_\Sigma\!\! d^3{\bf
x}\,\,\bar\alpha^-[\hat\partial^{-1}, {\cal A}] \alpha^-
\,,
\label{eq:lcqcd-massint2}
\\
L_{m,\phi\Lambda
\alpha}^{-+-}&=&-m\int_\Sigma\!\! d^3{\bf
x}\,\,\Big(\phi^-\Lambda\hat\partial^{-1}\alpha^--(\hat\partial^{-1}\bar\alpha^-)\Lambda
\bar\phi^-\Big)\,.
\label{eq:lcsqcd-massint2}
\end{eqnarray}

\section{Summary of the canonical transformation for MHV-QCD}
\label{sect:sum-canonical-QCD}
\begin{eqnarray}
{\cal A}_q^{\rm QCD} &=& \sum_{n=1}^\infty\int_{1\cdots
n}\!\!\!\Upsilon_{q,\bar1\cdots\bar n}{\cal B}_{1}\cdots
{\cal B}_{n}\:\delta_{q\bar1\cdots\bar n},
\\
\alpha ^{+,\rm QCD}_q&=&\sum_{n=1}^\infty\int_{1\cdots
n}\!\!\!\Upsilon^+_{q,\bar1\cdots\bar n}{\cal B}_{1}\cdots
{\cal B}_{n-1}\xi_n^+\:\delta_{q\bar1\cdots\bar n},
\\
\bar\alpha ^{+,\rm QCD}_q&=&\sum_{n=1}^\infty\int_{1\cdots
n}\!\!\!\Xi^+_{q,\bar1\cdots\bar n}\bar\xi^+_1{\cal B}_{2}\cdots
{\cal B}_{n}\:\delta_{q\bar1\cdots\bar n},
\\
\alpha^{-,\rm QCD}_q&=&\xi^-_q+\sum_{n=2}^\infty\int_{1\cdots
n}\!\!\!\Upsilon^-_{q,\bar1\cdots\bar n} {\cal B}_{1}\cdots
{\cal B}_{n-1}\xi_n^-\:\delta_{q\bar1\cdots\bar n},
\\
\bar\alpha^{-,\rm QCD}_q&=&\bar\xi^-_q+\sum_{n=2}^\infty\int_{1\cdots
n}\!\!\!\Xi^-_{q,\bar1\cdots\bar n}\bar\xi^+_1{\cal B}_{2}\cdots
{\cal B}_{n}\:\delta_{q\bar1\cdots\bar n},
\end{eqnarray}
\begin{eqnarray}
\bar {\cal A}^{\rm QCD}_q&=&\bar{\cal A}^{B,\rm QCD}+\bar{\cal
A}^{\xi B,\rm QCD}\,,
\label{eq:canon-trans-barA-1-QCD} \\
\bar{\cal A}^{B,\rm QCD}_q&=&\frac 1 {\hat q}\sum_{n=1}^\infty\int_{1\cdots
n}\sum_{s=1}^{n}\hat s\Xi^{s}_{q,\bar1\cdots\bar n}{\cal
B}_{1}\cdots\bar {\cal B}_s\cdots
{\cal B}_{n}\:\delta_{q\bar1\cdots\bar n}\,,
\\
\bar{\cal A}^{\xi B,\rm QCD}_q&=& \frac {g^2}{2\sqrt 2\hat q}\sum_{n=2}^\infty\int_{1\cdots
n}\bigg[\sum_{s=1}^{n-1}K^{+,s}_{q,\bar1\cdots\bar n}\, {\cal B}_{1}\cdots
{\cal B}_{s-1}\xi_s^+\bar\xi^-_{s+1}{\cal B}_{s+2}\cdots
{\cal B}_{n}
\nonumber 
\\
&&+\frac 1{N_C}K^{+,N_C}_{q,\bar1\cdots\bar n}\,\bar\xi^-_{1}{\cal B}_{2}\cdots
{\cal B}_{n-1}\xi^+_n{\bf I}
\nonumber\\
&&
+\sum_{s=1}^{n-1}K^{-,s}_{q,\bar1\cdots\bar n}\, {\cal B}_{1}\cdots
{\cal B}_{s-1}\xi^-_{s}\bar\xi^+_{s+1}{\cal B}_{s+2}\cdots
{\cal B}_{n}
\nonumber \\
&&+\frac 1{N_C}K^{-,N_C}_{q,\bar1\cdots\bar
n}\,\bar\xi^+_{1}{\cal B}_{2}\cdots
{\cal B}_{n-1}\xi^-_n{\bf I}\bigg]
\delta_{q\bar1\cdots\bar n}\,,
\end{eqnarray}
where
\begin{equation}
\delta_{q\bar1\cdots\bar n}=(2\pi)^3\delta^3(\vec q-\vec p_1-\cdots
\vec p_n)\,.
\end{equation} 
Define:
\begin{eqnarray}
\Delta_{\bar1\cdots\bar n}&=&
\begin{cases}
\frac {\hat 1\cdots \hat n}{\hat 1 \hat n(1\:2)\cdots (n-1,n)}
\,, \quad & \text{for } n\ge 2 \,,
 \cr
 \frac 1{\hat 1}\,, \quad &\text {for } n=1.
\end{cases}
\label{eq:Delta-def}
\end{eqnarray}
All the coefficients can be expressed as
\begin{eqnarray}
\Upsilon_{q,\bar1\cdots\bar n}&=&\Upsilon^+_{q,\bar1\cdots\bar
n}=\Xi^+_{q,\bar1\cdots\bar n}=(-i)^{n-1}\hat q\, \Delta_{\bar1\cdots\bar n}
\,,\\
\Xi^-_{q,\bar1\cdots\bar n}&=&(-i)^{n-1}\hat 1\, \Delta_{\bar1\cdots\bar n}
\,,\\
\Upsilon^-_{q,\bar1\cdots\bar n}&=&(-i)^{n-1}\hat n\,\Delta_{\bar1\cdots\bar n}
\,,\\
\Xi^{s}_{q,\bar1\cdots\bar n}&=&(-i)^{n-1}\hat s\, \Delta_{\bar1\cdots\bar n}
\,,\\
K^{+,s}_{q,\bar1\cdots\bar n}&=&(-i)^{n-1}\widehat {s+1}\, \Delta_{\bar1\cdots\bar n}
\,,\\
K^{+,N_C}_{q,\bar1\cdots\bar n}&=&-(-i)^{n-1}\hat {1}\, \Delta_{\bar1\cdots\bar n}
\,,\\
K^{-,s}_{q,\bar1\cdots\bar n}&=&-(-i)^{n-1}\hat {s}\, \Delta_{\bar1\cdots\bar n}
\,,\\
K^{-,N_C}_{q,\bar1\cdots\bar n}&=&(-i)^{n-1}\hat {n}\, \Delta_{\bar1\cdots\bar n}
\,.
\end{eqnarray}

\section{Summary of the canonical transformation for MHV-SQCD}
\label{sect:sum-canonical-SQCD}
\begin{eqnarray}
{\cal A}_q &=&{\cal A}^{\rm QCD}_q \,,
\quad
\alpha ^+_q=\alpha ^{+,\rm QCD}_q\,,
\quad
\bar\alpha ^+_q=\bar\alpha ^{+,\rm QCD}_q\,,
\\
\alpha^-_q&=&\alpha^{-,\rm QCD}_q-\sum_{n=2}^\infty\int_{1\cdots
n}\!\!\!\Upsilon^-_{q,\bar1\cdots\bar n}\Big( \sum_{l=1}^{n-1}{\cal
B}_{1}\cdots\Pi_l\cdots
{\cal B}_{n-1}\bar\varphi^-_n\Big)\delta_{q\bar1\cdots\bar n},
\\
\bar\alpha^-_q&=&\bar\alpha^{-,\rm QCD}_q+\sum_{n=2}^\infty\int_{1\cdots
n}\!\!\!\Xi^-_{q,\bar1\cdots\bar n}\Big(\sum_{l=2}^{n}\varphi^-_1{\cal
B}_{2}\cdots\Pi_l\cdots
{\cal B}_{n}\Big)\delta_{q\bar1\cdots\bar n},
\end{eqnarray}
\begin{eqnarray}
{\Lambda}_q&=&\sum_{n=1}^\infty\int_{1\cdots
n}\!\!\!\Upsilon_{q,\bar1\cdots\bar n}\sum_{l=1}^{n}{\cal
B}_{1}\cdots\Pi_l\cdots
{\cal B}_{n}\:\delta_{q\bar1\cdots\bar n},
\\
\phi^+_q&=&\varphi^+_q+\frac 1 {\hat q}\sum_{n=2}^\infty\int_{1\cdots
n}\!\!\!\Upsilon^+_{q,\bar1\cdots\bar n}\Big(\hat n {\cal B}_{1}\cdots
{\cal B}_{n-1}\varphi_n^+-\frac 1{ \sqrt 2}\sum_{l=1}^{n-1}{\cal
B}_{1}\cdots\Pi_l\cdots
{\cal B}_{n-1}\xi^+_n\Big)\delta_{q\bar1\cdots\bar n},
\\
\bar\phi^+_q&=&\bar\varphi^+_q+\frac 1 {\hat q}\sum_{n=2}^\infty\int_{1\cdots
n}\!\!\!\Xi^+_{q,\bar1\cdots\bar n}\Big(\hat 1 \bar\varphi^+_1{\cal B}_{2}\cdots
{\cal B}_{n}-\frac 1{ \sqrt 2}\sum_{l=2}^{n}\bar\xi^+_1{\cal
B}_{2}\cdots\Pi_l\cdots
{\cal B}_{n}\Big)\delta_{q\bar1\cdots\bar n},
\\
\bar\phi^-_q&=&\sum_{n=1}^\infty\int_{1\cdots
n}\!\!\!\Upsilon^-_{q,\bar1\cdots\bar n}{\cal B}_{1}\cdots
{\cal B}_{n-1}\bar\varphi_n^-\:\delta_{q\bar1\cdots\bar n},
\\
\phi ^-_q&=&\sum_{n=1}^\infty\int_{1\cdots
n}\!\!\!\Xi^-_{q,\bar1\cdots\bar n}\varphi^-_1{\cal B}_{2}\cdots
{\cal B}_{n}\:\delta_{q\bar1\cdots\bar n},
\end{eqnarray}

\begin{eqnarray}
\bar \Lambda_q&=&\bar\Lambda^{B\Pi}_q+\bar\Lambda^{\phi\xi B}_q\,,
\label{eq:canon-trans-barL-1}\\
\bar\Lambda^{B\Pi}_q&=&\sum_{n=1}^\infty\int_{1\cdots
n}\sum_{l=1}^{n}\Xi^{l}_{q,\bar1\cdots\bar n}{\cal
B}_{1}\cdots\bar\Pi_l\cdots
{\cal B}_{n}\:\delta_{q\bar1\cdots\bar n}\,,
\label{eq:canon-trans-barL-2} 
\\
\bar\Lambda^{\varphi\xi B}_q&=& \frac {g^2}2\sum_{n=2}^\infty\int_{1\cdots
n}\bigg[\sum_{s=1}^{n-1}K^{+,s}_{q,\bar1\cdots\bar n} {\cal B}_{1}\cdots
{\cal B}_{s-1}\xi_s^+\varphi^-_{s+1}{\cal B}_{s+2}\cdots
{\cal B}_{n}
\nonumber\\
&&-\frac 1{N_C}K^{+,N_C}_{q,\bar1\cdots\bar n}\varphi^-_{1}{\cal B}_{2}\cdots
{\cal B}_{n-1}\xi^+_n{\bf I}
\nonumber\\
&&
+\sum_{s=1}^{n-1}K^{-,s}_{q,\bar1\cdots\bar n} {\cal B}_{1}\cdots
{\cal B}_{s-1}\bar\varphi^-_{s}\bar\xi^+_{s+1}{\cal B}_{s+2}\cdots
{\cal B}_{n}
\nonumber \\
&&-\frac 1{N_C}K^{-,N_C}_{q,\bar1\cdots\bar
n}\bar\xi^+_{1}{\cal B}_{2}\cdots
{\cal B}_{n-1}\bar\varphi^-_n{\bf I}\bigg]
\delta_{q\bar1\cdots\bar n}
\,.\label{eq:canon-trans-barL-3}
\end{eqnarray}

\begin{eqnarray}
\bar {\cal A}_q&=&\bar{\cal A}^{B}+\bar{\cal A}^{B\Pi}+\bar{\cal
A}^{\xi B}+\bar{\cal A}^{\varphi B}+\bar{\cal A}^{\varphi\xi\Pi B}\,,
\label{eq:canon-trans-barA-1} 
\end{eqnarray}
\begin{eqnarray}
\bar{\cal A}^{B}_q&=&\bar{\cal A}^{B,\rm QCD}_q\,,
 \quad
\bar{\cal A}^{\xi B}_q= \bar{\cal A}^{\xi B,\rm QCD}_q\,,
\label{eq:canon-trans-barA-2}
\end{eqnarray}
\begin{eqnarray}
\bar{\cal A}^{B\Pi}_q&=&-\frac 1 {\sqrt2 \hat q}\sum_{n=2}^\infty\int_{1\cdots
n}\sum_{s=1}^{n}\bigg[\Xi^{s}_{q,\bar1\cdots\bar
n}\sum_{l=1,l\neq s}^{n}(-1)^{\delta_{ls}}{\cal
B}_{1}\cdots\Pi_l\cdots\bar \Pi_s\cdots
{\cal B}_{n}\bigg]\delta_{q\bar1\cdots\bar n}\,,
\label{eq:canon-trans-barA-3} 
\end{eqnarray}
\begin{eqnarray}
\bar{\cal A}^{\varphi B}_q&=& \frac {g^2}{2\hat q}\sum_{n=2}^\infty\int_{1\cdots
n}\bigg[\sum_{s=1}^{n-1}\hat s K^{+,s}_{q,\bar1\cdots\bar n}\, {\cal B}_{1}\cdots
{\cal B}_{s-1}\varphi_s^+\varphi^-_{s+1}{\cal B}_{s+2}\cdots
{\cal B}_{n}
\nonumber \\
&&-\frac{\hat n}{N_C}K^{+,N_C}_{q,\bar1\cdots\bar n}\,\varphi^-_{1}{\cal B}_{2}\cdots
{\cal B}_{n-1}\varphi^+_n{\bf I}
\nonumber\\
&&
-\sum_{s=1}^{n-1}\widehat{s+1}\,K^{-,s}_{q,\bar1\cdots\bar n}\, {\cal B}_{1}\cdots
{\cal B}_{s-1}\bar\varphi^-_{s}\bar\varphi^+_{s+1}{\cal B}_{s+2}\cdots
{\cal B}_{n}
\nonumber \\
&&+\frac {\hat 1}{N_C}K^{-,N_C}_{q,\bar1\cdots\bar
n}\,\bar\varphi^+_{1}{\cal B}_{2}\cdots
{\cal B}_{n-1}\bar\varphi^-_n{\bf I}\bigg]
\delta_{q\bar1\cdots\bar n}\,,
\label{eq:canon-trans-barA-4} 
\end{eqnarray}
\begin{eqnarray}
\bar{\cal A}^{\varphi \xi \Pi B}_q&=& \frac {g^2}{2\sqrt 2 \hat q}\sum_{n=3}^\infty\int_{1\cdots
n}\bigg[-\sum_{s=1}^{n-1} K^{+,s}_{q,\bar1\cdots\bar
n}\bigg(\sum_{l=1,l\ne s,s+1}^n (-1)^{\delta_{ls}}{\cal B}_{1}\cdots
{\Pi}_{l}\cdots\xi_s^+\varphi^-_{s+1}\cdots
{\cal B}_{n}\bigg)
\nonumber \\
&&+\frac{1}{N_C}K^{+,N_C}_{q,\bar1\cdots\bar
n}\sum_{s=2}^{n-1}\varphi^-_{1}{\cal B}_{2}\cdots
\Pi_s\cdots{\cal B}_{n-1}\xi^+_n{\bf I}
\nonumber\\
&&
-\sum_{s=1}^{n-1}K^{-,s}_{q,\bar1\cdots\bar n}\bigg(\sum_{l=1,l\ne s,s+1}^n
(-1)^{\delta_{ls}}{\cal B}_{1}\cdots
{\Pi}_{l}\cdots\bar\varphi^-_{s}\bar\xi^+_{s+1}{\cal B}_{s+2}\cdots
{\cal B}_{n}\bigg)
\nonumber\\
&&-\frac {1}{N_C}K^{-,N_C}_{q,\bar1\cdots\bar
n}\sum_{s=2}^{n-1}\bar\xi^+_{1}{\cal B}_{2}\cdots\Pi_s\cdots
{\cal B}_{n-1}\bar\varphi^-_n{\bf I}\bigg]
\delta_{q\bar1\cdots\bar n}\,,
\label{eq:canon-trans-barA-5} 
\end{eqnarray}
where $\bf I$ is the color singlet unit matrix and 
\begin{eqnarray}
\delta_{ls}=
\left\{ 
\begin{array}{ll}
0& \text{ for }
l<s \,,\cr
1 & \text { for } l>s\,.
\end{array}
\right.
\label{eq:delta-ls}
\end{eqnarray}

\section{Summary of the massive CSW vertices for MHV-QCD}
\label{sect:mass-V-QCD}

\begin{eqnarray}
 L_m^{+-}&=&\frac {m^2}{\sqrt 2}\bigg[
\sum_{n=2}^\infty\int_{1,2,\cdots n}V_m^{-+}(\bar1\cdots \bar n)\bar
\xi_1^-{\cal B}_2\cdots{\cal B}_{n-1}\xi^+_n\:\delta_{1\cdots n}
\nonumber\\
&&+ \sum_{n=2}^\infty\int_{1,2,\cdots n}V_m^{+-}(\bar 1\cdots \bar n)\bar
\xi_1^+{\cal B}_2\cdots{\cal B}_{n-1}\xi^-_n\:\delta_{1\cdots n}
\bigg]\,,
\label{eq:Lpm-mass-expansion}
\end{eqnarray}
where 
\begin{eqnarray}
\delta_{1\dots n}&=&(2\pi)^3 \delta^3\Big(\sum_{i=1}^n p_i\Big)\,,
\label{eq:delta-1n}
\\
V_m^{-+}(\bar 1\cdots\bar n)&=&(-i)^{n-2}\Delta_{\bar1\cdots\bar n} \frac{(1\:n)}
{\hat n}\,,
\\
V_m^{+-}(\bar 1\cdots \bar n)&=&-(-i)^{n-2}\Delta_{\bar1\cdots\bar n} \frac{(1\:n)}
{\hat 1}\,,
\end{eqnarray}
and $\Delta$ is defined in (\ref{eq:Delta-def}).

\begin{eqnarray}
L^{+-+}_{Fm}&=&im\bigg[
\sum_{n=3}^\infty\sum_{s=2}^{n-1}\int_{1\cdots n}V^{s,+-+}(\bar
1\cdots\bar n) \bar\xi_1^+ {\cal
B}_2\cdots\bar{\cal B}_s\cdots{\cal B}_{n-1}\xi^+_n\:\delta_{1\cdots n}
\nonumber \\
&&\hspace{-1cm}+\frac{g^2}{2\sqrt2}\sum_{n=4}^\infty\sum_{s=2}^{n-2}\int_{1,2,\cdots n}
\bigg(V^{s,++-+}(\bar 1\cdots\bar  n) \bar\xi_1^+ {\cal
B}_2\cdots{\cal B}_{s-1}\xi_s^+\bar\xi^-_{s+1}
{\cal B}_{s+2}\cdots{\cal B}_{n-1}\xi^+_n
\nonumber \\
&&+V^{s,+-++}(\bar 1\cdots \bar n) \bar\xi_1^+ {\cal
B}_2\cdots{\cal B}_{s-1}\xi_s^-\bar\xi^+_{s+1}
{\cal B}_{s+2}\cdots{\cal B}_{n-1}\xi^+_n\bigg)\delta_{1\cdots n}\bigg
]\,,
\label{eq:Lpmp-mass-expansion}
\end{eqnarray}
where 
\begin{eqnarray} 
V^{s,+-+}(\bar 1\cdots\bar  n)&=&(-i)^{n-3}\Delta_{\bar1\cdots\bar n}\frac{(1\:s)(s\:n)}{\hat1 \hat
n}\,, \quad \text{for } n\geq3\,,
\nonumber \\
V^{s,++-+}(\bar 1\cdots\bar n)&=&(-i)^{n-3}\Delta_{\bar1\cdots\bar n}\frac{(1\:s)}{
\hat1\hat s}\bigg( \frac{(s+1\:n)}{  \hat n }-\frac 1
{N_C} \frac {(s\:s+1)}{ \hat s  }\bigg)
\,,\quad \text{for } n\geq 4\,,
\nonumber \\
V^{s,+-++}(\bar 1\cdots\bar n)&=&-(-i)^{n-3}\Delta_{\bar1\cdots\bar
n}\frac{(s+1\:n)}{\widehat{ s+1} \hat n}
\bigg( \frac{(1\:s)}{ \hat1}-\frac 1
{N_C} \frac {(s\:s+1)}{ \widehat{ s+1} }\bigg)
\,, \quad \text{for } n\geq4\,.
\end{eqnarray}
\begin{eqnarray}
L^{-+-}_{Fm}=im\sum_{n=3}^\infty\int_{1,2,\cdots n}
V^{-+-}(\bar1\cdots\bar n) \bar\xi_1^- {\cal
B}_2{\cal B}_3\cdots{\cal B}_{n-1}\xi^-_n\:\delta_{1\cdots n}\,,
\label{eq:Lmpm-mass-expansion}
\end{eqnarray}
where 
\begin{eqnarray}
V^{-+-}(\bar1\cdots\bar n)=-(-i)^{n-3}\Delta_{\bar1\cdots\bar n}\frac{(1\:n)^2}{\hat1 \hat
n}\,, \quad \text{for } n\geq3\,. 
\end{eqnarray}


\begin{thebibliography}{99}
\bibitem{Parke:1986gb} S.~J.~Parke and T.~R.~Taylor,
  Phys.\ Rev.\ Lett.\ {\bf 56} (1986) 2459.

\bibitem{Berends:1988zn} F.~A.~Berends and W.~T.~Giele,
  Nucl.\ Phys.\ B {\bf 313}, 595 (1989).

\bibitem{Cachazo:2004kj} F.~Cachazo, P.~Svr\v cek and E.~Witten,
  JHEP {\bf 0409}, 006 (2004) [arXiv:hep-th/0403047].

\bibitem{Wu:2004fb} J.~B.~Wu and C.~J.~Zhu,
  JHEP {\bf 0407}, 032 (2004) [arXiv:hep-th/0406085];

\bibitem{Wu:2004jxa} J.~B.~Wu and C.~J.~Zhu,
  JHEP {\bf 0409} (2004) 063 [arXiv:hep-th/0406146].

\bibitem{Georgiou:2004wu} G.~Georgiou and V.~V.~Khoze,
  JHEP {\bf 0405}, 070 (2004) [arXiv:hep-th/0404072];
\bibitem{Georgiou:2004by} G.~Georgiou, E.~W.~N.~Glover and
  V.~V.~Khoze,
  JHEP {\bf 0407}, 048 (2004) [arXiv:hep-th/0407027].

\bibitem{Dixon:2004za}
  L.~J.~Dixon, E.~W.~N.~Glover and V.~V.~Khoze,
  JHEP {\bf 0412} (2004) 015
  [arXiv:hep-th/0411092].

\bibitem{Badger:2004ty}
  S.~D.~Badger, E.~W.~N.~Glover and V.~V.~Khoze,
  JHEP {\bf 0503} (2005) 023
  [arXiv:hep-th/0412275].

\bibitem{Bern:2004ba}
  Z.~Bern, D.~Forde, D.~A.~Kosower and P.~Mastrolia,
  Phys.\ Rev.\  D {\bf 72}, 025006 (2005)
  [arXiv:hep-ph/0412167].


\bibitem{Brandhuber:2004yw} A.~Brandhuber, B.~J.~Spence and
  G.~Travaglini,
  Nucl.\ Phys.\ B {\bf 706} (2005) 150 [arXiv:hep-th/0407214].

\bibitem{Bedford:2004py} J.~Bedford, A.~Brandhuber, B.~J.~Spence and
  G.~Travaglini,
  Nucl.\ Phys.\ B {\bf 706} (2005) 100 [arXiv:hep-th/0410280].

\bibitem{Bedford:2004nh} J.~Bedford, A.~Brandhuber, B.~J.~Spence and
  G.~Travaglini,
  Nucl.\ Phys.\ B {\bf 712} (2005) 59 [arXiv:hep-th/0412108].

\bibitem{Brandhuber:2005kd} A.~Brandhuber, B.~Spence and
  G.~Travaglini,
  JHEP {\bf 0601} (2006) 142 [arXiv:hep-th/0510253].

\bibitem{Quigley:2004pw} C.~Quigley and M.~Rozali,
  JHEP {\bf 0501} (2005) 053 [arXiv:hep-th/0410278].


\bibitem{Brandhuber:2006bf} A.~Brandhuber, B.~Spence and
  G.~Travaglini,
  JHEP {\bf 0702} (2007) 088 [arXiv:hep-th/0612007].

\bibitem{Brandhuber:2007vm} A.~Brandhuber, B.~Spence, G.~Travaglini
  and K.~Zoubos,
  JHEP {\bf 0707} (2007) 002 [arXiv:0704.0245 [hep-th]].

\bibitem{Britto:2005fq}
  R.~Britto, F.~Cachazo, B.~Feng and E.~Witten,
  Phys.\ Rev.\ Lett.\  {\bf 94} (2005) 181602
  [arXiv:hep-th/0501052].

 
\bibitem{Risager:2005vk} K.~Risager,
  JHEP {\bf 0512} (2005) 003 [arXiv:hep-th/0508206].

\bibitem{Britto:2004ap} R.~Britto, F.~Cachazo and B.~Feng,
  Nucl.\ Phys.\ B {\bf 715} (2005) 499 [arXiv:hep-th/0412308];

\bibitem{Gorsky:2005sf} A.~Gorsky and A.~Rosly,
  JHEP {\bf 0601} (2006) 101 [arXiv:hep-th/0510111].


\bibitem{Mansfield:2005yd} P.~Mansfield,
  JHEP {\bf 0603}, 037 (2006) [arXiv:hep-th/0511264].

\bibitem{Ettle:2006bw} J.~H.~Ettle and T.~R.~Morris,
  JHEP {\bf 0608}, 003 (2006) [arXiv:hep-th/0605121].

\bibitem{Ettle:2007qc} J.~H.~Ettle, C.~H.~Fu, J.~P.~Fudger,
  P.~R.~W.~Mansfield and T.~R.~Morris,
  JHEP {\bf 0705}, 011 (2007) [arXiv:hep-th/0703286].

\bibitem{Boels:2007qn}
  R.~Boels, L.~Mason and D.~Skinner,
  Phys.\ Lett.\  B {\bf 648} (2007) 90
  [arXiv:hep-th/0702035].

\bibitem{Boels:2006ir}
  R.~Boels, L.~Mason and D.~Skinner,
  JHEP {\bf 0702} (2007) 014
  [arXiv:hep-th/0604040].

\bibitem{Boels:2007pj} R.~Boels and C.~Schwinn,
  Phys.\ Lett.\ B {\bf 662} (2008) 80 [arXiv:0712.3409 [hep-th]].

\bibitem{Boels:2008ef} R.~Boels and C.~Schwinn,
  arXiv:0805.1197 [hep-th].

\bibitem{Ettle:2008ey}
  J.~H.~Ettle, T.~R.~Morris and Z.~Xiao,
  JHEP {\bf 0808} (2008) 103
  [arXiv:0805.0239 [hep-th]].

\bibitem{Boels:2008du}
  R.~Boels and C.~Schwinn,
  arXiv:0805.4577 [hep-th].


\bibitem{Schwinn:2008fm}
  C.~Schwinn,
  arXiv:0809.1442 [hep-ph].

\bibitem{Schwinn:2006ca}
  C.~Schwinn and S.~Weinzierl,
  JHEP {\bf 0603} (2006) 030
  [arXiv:hep-th/0602012].

\bibitem{Feng:2006yy} H.~Feng and Y.~t.~Huang,
  arXiv:hep-th/0611164.

\end{thebibliography}
\end{document}